\documentclass[aps,prd,amsmath,amssymb,showpacs,showkeys,eqsecnum,nofootinbib]{revtex4}
\usepackage{graphicx}
\usepackage{dcolumn}
\usepackage{bm}
\usepackage{epsfig}

\begin{document}
\title{Bound-state parameters from dispersive sum rules for vacuum-to-vacuum correlators}
\author{Wolfgang Lucha$^{a}$, Dmitri Melikhov$^{a,b,c}$, and Silvano Simula$^{d}$} 
\affiliation{
$^a$Institute for High Energy Physics,
Austrian Academy of Sciences, Nikolsdorfergasse 18, A-1050, Vienna, Austria\\
$^b$Faculty of Physics, University of Vienna, Boltzmanngasse 5, A-1090 Vienna, Austria\\
$^c$SINP, Moscow State University, 119991, Moscow, Russia\\
$^d$INFN, Sezione di Roma III, Via della Vasca Navale 84, I-00146, Roma, Italy}
\date{\today}
\begin{abstract}
We study the extraction of the ground-state parameters from vacuum-to-vacuum correlators.  
We work in quantum-mechanical potential model which provides the only possibility to probe 
the reliability and the actual accuracy of this method: one obtains the bound-state parameters 
from the correlators by the standard procedures adopted in the method of sum rules 
and compares these results with the exact values calculated from the Schr\"odinger equation.  
We focus on the crucial ingredient of the method of sum rules 
--- the effective continuum threshold --- and propose a new algorithm to fix this quantity. 
In a quantum-mechanical model, our procedure leads to a remarkable improvement 
of the accuracy of the extracted ground-state parameters compared to the standard procedures adopted 
in the method and used in all previous applications of dispersive sum rules in QCD. 
The application of the proposed procedure in QCD promises a considerable increase of the accuracy 
of the extracted hadron parameters. 
\end{abstract}
\pacs{11.55.Hx, 12.38.Lg, 03.65.Ge}
\keywords{Nonperturbative QCD, hadron properties, QCD sum rules}
\maketitle

\section{Introduction}
%

The extraction of the ground-state parameters from the operator product expansion (OPE) series for a relevant correlator 
is a cumbersome procedure: even if several terms of 
the OPE for the correlator are known precisely, the numerical procedures of the method 
of dispersive sum rules \cite{svz,ioffe} cannot determine the true exact value of the bound-state parameter. 
Instead, the method should provide the band of values such the true hadron parameter has a 
flat probability distribution within this band \cite{svz}. This band is a systematic, or intrinsic, 
sum-rule uncertainty. 
 
The method of sum rules in QCD contains a set of prescriptions (see e.g.~\cite{lms_lcsr,lms_2ptsr}) 
which are believed to provide such a systematic error. In QCD this, however, always remains a conjecture ---  
it is impossible to prove that the range provided by the standard sum-rule  
procedures indeed contains the actual value of the bound-state parameter. 

The only possibility to acquire an unbiased judgement of the reliability of the error
estimates in sum rules is to apply the method to a problem where the parameters 
of the theory may be fixed and the corresponding parameters of the ground state 
may be calculated independently and exactly. Presently, only quantum-mechanical potential models 
provide such a possibility. 

A simple harmonic-oscillator (HO) potential model 
\cite{lms_2ptsr,lms_3ptsr,m_lcsr,lms_prl} 
possesses the essential features of QCD --- 
confinement and asymptotic freedom \cite{nsvz} --- and has 
the following advantages: 
(i) the bound-state parameters (masses, wave functions, form factors) are known precisely;
(ii) direct analogues of the QCD correlators may be calculated exactly. 
(For a discussion of many aspects of sum rules in quantum mechanics we refer to 
\cite{nsvz1,qmsr,orsay,ms_inclusive,radyushkin2001,bakulev}).

Making use of this model, we have already studied the extraction of ground-state parameters from 
different types of correlators: namely, the ground-state decay 
constant from two-point vacuum-to-vacuum correlator \cite{lms_2ptsr}, the form factor from three-point 
vacuum-to-vacuum correlator 
\cite{lms_3ptsr}, and form factor from vacuum-to-hadron correlator \cite{m_lcsr}. We have demonstrated 
that the standard adopted procedures for obtaining the systematic errors do not work properly: 
for all types of correlators the true known value of the bound-state parameter was shown to lie outside 
the band obtained according to the standard criteria. 
These results give a solid ground to claim that also in QCD the actual accuracy of the method turns out to be 
much worse than the accuracy expected on the basis of the standard criteria applied. 

Thus, our results reported in \cite{lms_2ptsr,lms_3ptsr,m_lcsr,lms_prl} mainly contained cautious messages 
concerning the application of sum rules to hadron properties. 
However, understanding the problem is the necessary first step in solving the problem. 

We have realized that the main origin of this failure of the method lies in an over-simplified model 
for hadron contnuum which is described as a perturbative contribution above a constant Borel-parameter 
independent effective continuum threshold. 
We have introduced the notion of the {\it exact} effective continuum threshold, 
which corresponds to the true bound-state 
parameters: in a HO model the true hadron parameters --- decay constant and form factor --- 
are known and the exact effective continuum thresholds for different correlators may be calculated. We have demonstrated that 
the exact effective 
continuum threhsold (i) is not a universal quantity and depends on the correlator considered (i.e. it is in general 
different for two-point and three-point vacuum-to-vacuum correlators), and (ii) depends on the 
Borel parameter and, for the form-factor case, also on the momentum-transfer. 

Moreover, we have shown that the ``Borel stability criterium'' combined with the assumption of a 
Borel-parameter independent effective continuum threshold leads to the extraction of the wrong bound-state parameters. 

In view of these results a natural question arises: it is possible to formulate modified  
procedures which would improve the actual accuracy of hadron parameters extracted from sum rules, 
or the method is only suitable for obtaining numerical estimates with uncontrollable errors 
which may amount to 20-30\%? 

We believe the answer to the first part of this question is positive. 

In a recent paper \cite{lms_prl} we proposed a new algorithm for extracting the parameters of the ground state.
The simple idea of \cite{lms_prl} is to relax the standard assumption of a Borel-parameter 
independent effective continuum threshold, and allow for a Borel-parameter dependent quantity. 
In \cite{lms_prl} we have illustrated the way the procedure works for the extraction of the form factor 
from three-point vacuum-to-vacuum correlator.  

In the present paper we develop this idea and provide details of its application to the extraction 
of hadron parameters from dispersive sum rules. We present for the first time the application of the new 
algorithm to the two-point correlator and present a detailed analysis of the extraction of the form factor 
from three-point correlator. The new algorithm is shown to lead to a significant improvement of the actual accuracy 
of the extracted bound-state parameters. 
Even though the full control over the systematic uncertainties, in the strict mathematical sense, is still not feasible 
(and presumably cannot be achieved in principle), our findings suggest that it may be 
possible to shed light on the actual accuracy of the method of sum-rules in QCD.

The paper is organized as follows: Section \ref{Sect:HO} recalls the results for bound states in the HO model; 
Section \ref{Sect:Pi} gives the OPE for the polarization operator. 
Section \ref{Sect:Gamma} presents detailed results for the vertex function and the corresponding OPE.  
Sum rules for the decay constant and the form factor are given in Section \ref{Sect:SR}. 
Section \ref{Sect:Numerics} provides our numerical results. 
Finally, Section \ref{Sect:Conclusions} presents our concluding remarks.  

\section{\label{Sect:HO}Harmonic-oscillator model}
We consider a nonrelativistic model Hamiltonian $H$ with a HO
interaction potential $V(r)$, $r\equiv|\vec r\,|$:
\begin{eqnarray}
H=H_0+V(r), \qquad H_0={\vec p}^{\,2}/2m, \qquad V(r)={m\omega^2r^2}/{2}.
\end{eqnarray}
The full Green function $G(E)\equiv(H-E)^{-1}$ and the free Green
function $G_0(E)\equiv(H_0-E)^{-1}$ are related by
\begin{eqnarray}
G^{-1}(E)-G_0^{-1}(E)=V.
\end{eqnarray}
The solution $G(E)$ of this relation may be easily obtained as an expansion in powers of the interaction $V$:
\begin{eqnarray}
\label{ls} 
G(E)=G_0(E)-G_0(E)VG_0(E)+\cdots.
\end{eqnarray}
In the HO model, all characteristics of the bound states are
easily calculable. For instance, for the ground state (g), $n=0$, one finds 
\begin{eqnarray}
\label{E0} 
E_{\rm g}=\frac{3}{2}\omega,\qquad R_{\rm g}\equiv |\psi_{\rm g}(\vec r=0)|^2=
\left(\frac{m\omega}{\pi}\right)^{3/2},\qquad F_{\rm g}(q)=\exp(-q^2/4m\omega),
\end{eqnarray}
where the elastic form factor of the ground state is defined in terms of the ground-state wave 
function $\psi_{\rm g}$ according to
\begin{eqnarray}
\label{FG}
F_{\rm g}(q)=\langle \Psi_{\rm g}|J(\vec q)|\Psi_{\rm g}\rangle=
\int d^3k\,\psi^\dagger_{\rm g}(\vec k)\psi_{\rm g}(\vec k-\vec q)= \int d^3r\, |\psi_{\rm g}(\vec r)|^2e^{i\vec q\vec r},
\qquad  q\equiv |\vec q|, 
\end{eqnarray}
and the current operator $J(\vec q)$ is given by the kernel
\begin{eqnarray}
\label{J} \langle \vec r\,'|J(\vec q)|\vec r\rangle=
\exp(i\vec q\vec r)\delta^{(3)}(\vec r-\vec r\,').
\end{eqnarray}

\section{\label{Sect:Pi}Polarization operator}
The quantum-mechanical analogue of the polarization operator has the form 
\begin{eqnarray}
\label{pi} 
\Pi(T)=\langle \vec r_f=0|\exp(- H T)|\vec r_i=0\rangle. 
\end{eqnarray}
This quantity is used in the sum-rule approach for the extraction of the wave
function at the origin (i.e., of the decay constant) of the ground
state \cite{svz}. A detailed analysis of the corresponding procedure for the HO model can be found in \cite{lms_2ptsr}. 
For the HO potential, the analytic expression for $\Pi(T)$ is well-known \cite{nsvz}:
\begin{eqnarray}
\label{piexact} 
\Pi(T)=\left(\frac{\omega m}{\pi}\right)^{3/2}
\frac1{\left[2\sinh(\omega T)\right]^{3/2}}.
\end{eqnarray}
Apart from the overall factor, $\Pi(T)$ is a function of one parameter $T\omega$. 

The average energy of the polarization function is defined as 
\begin{eqnarray}
\label{energypi}
E_\Pi(T)\equiv -\partial_T\log \Pi(T)=\frac32 \omega \coth(\omega T),\qquad 
\partial_T\equiv \frac{\partial}{\partial T}.
\end{eqnarray}
At $T=0$ both $\Pi(T)$ and $E_\Pi(T)$ diverge. For large values of $T$ the contributions of the 
excited states to the correlator vanish and therefore 
$E_\Pi(T)\to E_{\rm g}$ for $T\to\infty$. The deviation of the energy from $E_{\rm g}$ at finite values of $T$
measures the ``contamination'' of the correlator  by the excited states. 

The ground-state contribution to the correlator has a simple form: 
\begin{eqnarray}
\label{piground} 
\Pi_{\rm g}(T)=\left(\frac{\omega m}{\pi}\right)^{3/2}
\exp\left({-\frac32\omega T}\right).
\end{eqnarray}
The OPE series is the expansion of $\Pi(T)$ at small Euclidean time $T$: 
\begin{eqnarray}
\label{piope} 
\Pi_{\rm OPE}(T)=\left(\frac{m}{2\pi T}\right)^{3/2}
\left(1-\frac{1}{4}\omega^2T^2+\frac{19}{480}{\omega^4 T^4}
+\cdots \right).
\end{eqnarray}
For $\Pi(T)$ this expansion is equivalent to the expansion in powers of the interaction $\omega$. 

The first term in (\ref{piope}) does not depend on the interaction and describes the free propagation 
of the constituents. $\Pi_0$ may be written as the spectral integral \cite{lms_2ptsr} 
\begin{eqnarray}
\label{pi0} 
\Pi_0(T)=\int dz e^{-z T} \rho_0(z), \qquad \rho(z)=\frac{2}{\sqrt{\pi}}\sqrt{z}, 
\end{eqnarray}
with $\rho_0(z)$ the known spectral density of the one-loop two-point Feynman diagram of the nonrelativistic 
field theory \cite{lms_2ptsr}. The rest of the series represents power corrections, which may be obtained 
just as the difference between the exact correlator and the free-propagation term: 
\begin{eqnarray}
\label{pipower} 
\Pi_{\rm power}(T)=\Pi(T)-\Pi_{0}(T).  
\end{eqnarray}

\section{\label{Sect:Gamma}Vertex function}
The basic quantity for the extraction of the form factor 
in the method of dispersive sum rules is the correlator of three currents \cite{ioffe}. 
The analogue of this quantity in quantum mechanics reads \cite{lms_3ptsr,radyushkin2001}
\begin{eqnarray}
\Gamma(E_2,E_1,q)= \langle \vec r_f=0|(H-E_2)^{-1}J(\vec q)(H-E_1)^{-1}|
\vec r_i=0\rangle,\qquad q\equiv |\vec q|,
\end{eqnarray}
[with the operator $J(\vec q)$ defined in (\ref{J})] and its
double Borel (Laplace) transform under $E_1\to \tau_1$ and $E_2\to\tau_2$
\begin{eqnarray}
\Gamma(\tau_2,\tau_1,q)= 
\langle \vec r_f=0|G(\tau_2)J(\vec q)G(\tau_1)|\vec r_i=0\rangle, \qquad 
G(\tau)\equiv \exp(-H \tau). 
\end{eqnarray}
For large $\tau_1$ and $\tau_2$ the correlator is dominated by the ground state:
\begin{eqnarray}
\Gamma(\tau_2,\tau_1,q)\to |\psi_{\rm g}(\vec r=0)|^2 e^{-E_{\rm g}(\tau_1+\tau_2)}F_{\rm g}(q^2)+\cdots.
\end{eqnarray}
Let us notice the Ward identity which relates the vertex function
at zero momentum to the polarization operator:
\begin{eqnarray}
\label{pigamma}
\Gamma(\tau_2,\tau_1,q=0)=\Pi(\tau_1+\tau_2).
\end{eqnarray}
This expression follows directly from the current conservation relation
\begin{eqnarray}
J(\vec q=0)=1.
\end{eqnarray}

\subsection{Exact $\Gamma(T,q)$ in the HO model}
We obtain the exact $\Gamma(\tau_2,\tau_1,q)$ by using the 
known expression for the Green function in configuration space 
\begin{eqnarray}
\label{gammar} 
\langle \vec r\,'=0|G(\tau)|\vec r\rangle=\Pi(\tau)\exp\left(-\frac{m\omega r^2}{2}\tanh^{-1}(\omega\tau)\right).
\end{eqnarray}
All necessary integrals are Gaussian and we easily derive an explicit expression for $\Gamma(\tau_2,\tau_1,q)$. 
For our further investigation, we consider the vertex function for equal times 
$\tau_1=\tau_2=\frac12 T$, which reads
\begin{eqnarray}
\label{gamma} 
\Gamma(T,q)=\Pi(T)
\exp\left(-\frac{q^2}{4m\omega}
\tanh\left(\frac{\omega T}{2}\right)\right).
\end{eqnarray}
The correlator $\Gamma$ is a function of two dimensionless variables $\omega T$ and $q^2/m\omega$. 
Setting, without loss of generality, $m=\omega$ 
we still have 2 dimensionless variables $\omega T$ and $q/\omega$. 
This will be done later for the numerical analysis. 

The corresponding average energy is defined as follows 
\begin{eqnarray}
\label{energygamma}
E_\Gamma(T,q)\equiv -\partial_T\log \Gamma(T,q)=\frac32 \omega \coth(\omega T)
+\frac{q^2}{4m}\frac{1}{\left(1+\cosh(\omega T)\right)}. 
\end{eqnarray}
\begin{figure}[!t]
\begin{tabular}{cc}
\includegraphics[width=6cm]{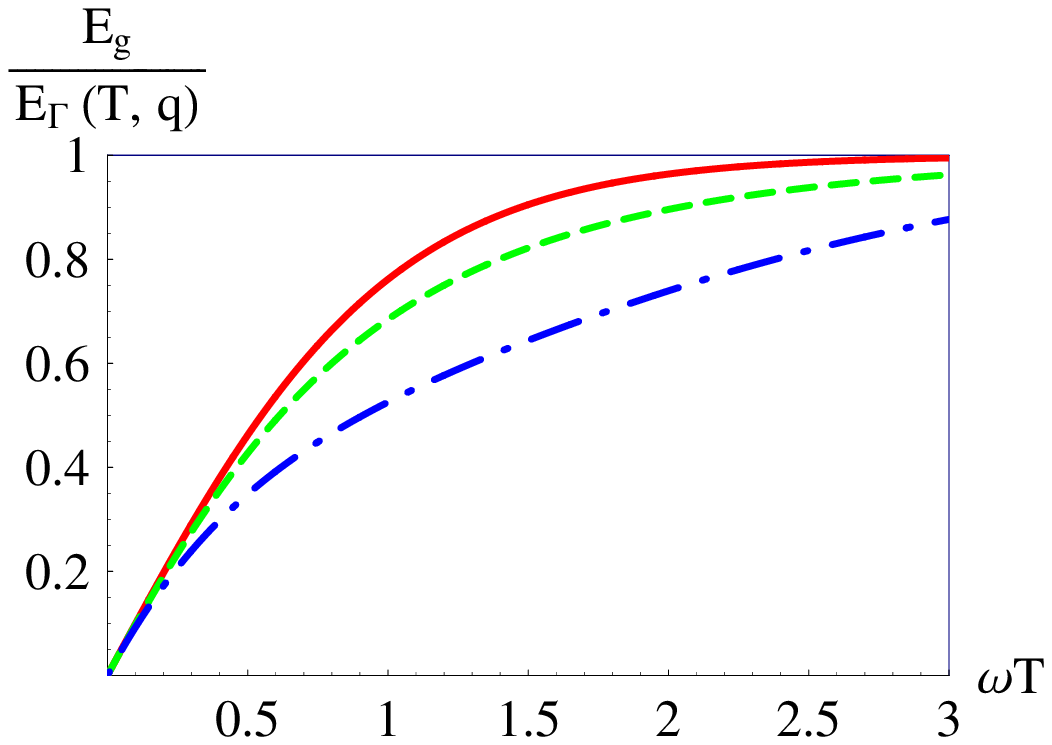}&
\includegraphics[width=6cm]{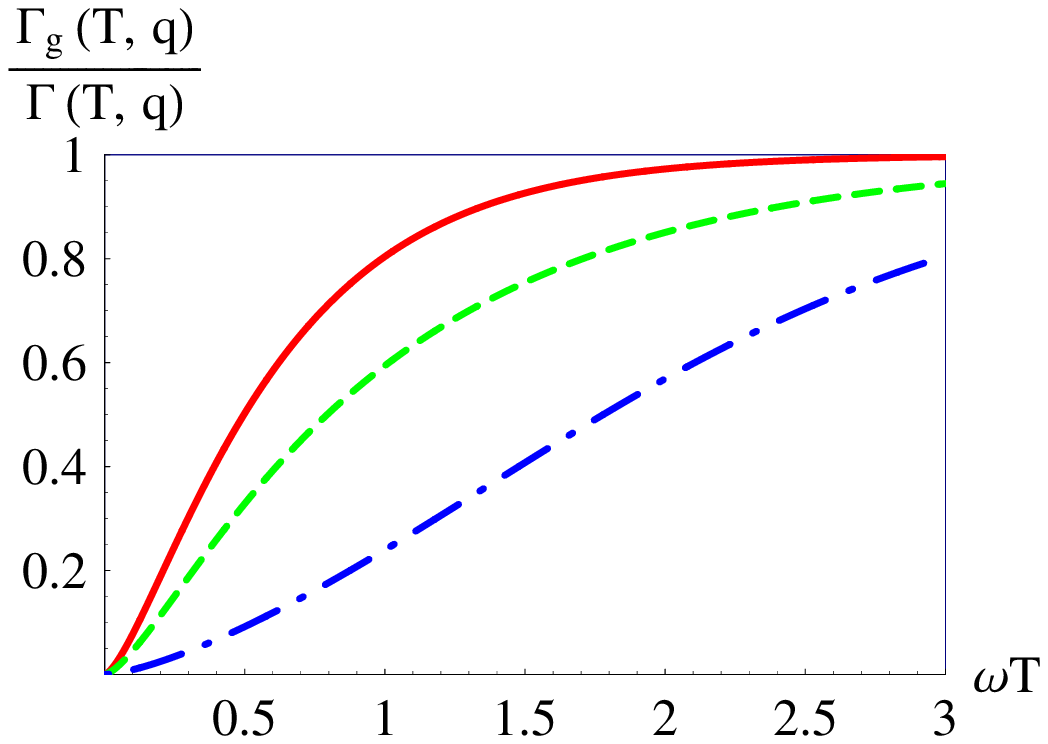}\\
(a)& (b)
\end{tabular}
\caption{\label{Plot:1} Relative ground-state contribution to the energy $E_{\rm g}(T,q)/E_\Gamma(T,q)$ (a) 
and to the correlator $\Gamma_{\rm g}(T,q)/\Gamma(T,q)$ (b) vs. $T$ for several values of $q$.  
Solid (red) line: $q=0$; dashed (green) line: $q=1.5\sqrt{m\omega}$; dash-dotted (blue) line:  
$q=3\sqrt{m\omega}$.}
\end{figure}
The contribution of the ground state to the correlator, $\Gamma_{\rm g}$, in the HO model has the form 
\begin{eqnarray}
\label{gammaground} 
\Gamma_{\rm g}(T,q)=\left(\frac{m\omega}{\pi}\right)^{3/2}
\exp\left({-\frac32\omega T}\right)
\exp\left(-\frac{q^2}{4m\omega}\right). 
\end{eqnarray}
The average energy $E_\Gamma$ and the ground-state contribution $\Gamma_{\rm g}$ vs. $T$ for different values of $q$ 
are shown in Fig.~\ref{Plot:1}.

First, notice that the on-set of the ground-state dominance in the correlator 
is shifted to later times with growing $q$. Second, if the ground-state energy $E_{\rm g}$ 
is known (hereafter we will be discussing 
precisely this situation), the deviation of the average correlator energy from the ground-state energy may be used 
as an indicator of the ground-state contribution to the correlator. This indicator is, however, not very accurate: 
the relative contribution of the ground state to the correlator turns out to be systematically greater 
than the deviation of the average correlator energy from the  ground-state energy. 

Figure \ref{Plot:1} makes obvious the way in which the ground-state form
factor may be extracted from the correlator $\Gamma(T,q)$ known
numerically (e.g., from the lattice): The correlator is dominated
by the ground state at large values of $T$; so one may calculate
the $T$- and $q$-dependent energy $E_{\Gamma}(T,q)$ (\ref{energygamma}) 
which exhibits a plateau at large $T$: $E(T,q)\to E_{\rm g}$ for any $q$. 
Making sure that one has already reached the plateau and that
the correlator is saturated by the ground state, one obtains the
form factor from the relation
\begin{eqnarray}
\label{plateau2}
F_{\rm g}(q)=\lim_{T\to\infty}\frac{1}{R_{\rm g}}e^{E_{\rm g}T}\Gamma(T,q).
\end{eqnarray}

\subsection{OPE for $\Gamma(T,q)$}
Let us now construct for $\Gamma(T,q)$ the analogue of the OPE
as used in the method of three-point sum rules in QCD. The
corresponding procedure consists of two steps: First, we expand
$\Gamma$ in powers of $\omega^2$ and obtain
\begin{eqnarray}
\label{gamma1}
\Gamma(T,q)=\sum\limits_{n=0}^\infty\Gamma_{2n}(q,T)\omega^{2n}.
\end{eqnarray}
Explicitly, for the lowest terms one has
\begin{eqnarray}
\label{gamma11}
\Gamma_0(T,q)&=&\left(\frac{m}{2\pi T}\right)^{3/2}\exp\left(-\frac{q^2T}{8m}\right),\nonumber\\
\Gamma_2(T,q)&=&\left(\frac{m}{2\pi T}\right)^{3/2}\exp\left(-\frac{q^2T}{8m}\right)
\left(-\frac14+\frac{q^2T}{96m}\right)\omega^2T^2. 
\end{eqnarray}
The first term, $\Gamma_0$, corresponds to free propagation and does not depend on the confining potential. 
It may be written as the double spectral representation \cite{lms_prd75}
\begin{eqnarray}
\Gamma_0(T,q)=\int dz_1 dz_2 e^{-{\frac12} z_1 T} e^{-{\frac12}
z_2 T}\Delta_0(z_1,z_2,q), \quad
\Delta_0(z_1,z_2,q)=\frac{m^2}{4\pi^2 q}\theta\left(\left(z_1+z_2-\frac{q^2}{2m}\right)^2-4z_1z_2<0\right).
\end{eqnarray}
The Ward identity in nonrelativistic field theory relates to each other the three-point function at 
zero momentum transfer and the two-point function and leads to 
\begin{eqnarray}
\label{wi}
\lim_{q\to 0}\Delta_0(z,z',q)=\delta(z-z')\rho_0(z). 
\end{eqnarray}
For the calculation of higher-order terms $\Gamma_{2n}$, $n\ge 1$, the explicit form of the confining 
potential is necessary. 
In the HO model it makes no problem to calculate all higher-order terms explicitly.   
Recall, however, that in QCD the situation is different: the confining potential is not known and therefore 
nonperturbative long-distance effects are parameterized as power corrections in terms of local condensates. 
Thus, for each quantity $\Gamma_{2n}$, $n\ge 1,$ one has a power-series expansion in $T$. To keep
the same track, we expand $\Gamma_{2n}$, $n\ge 1$, in powers of $T$.
As the result of this procedure, the quantum-mechanical analogue of the OPE for $\Gamma$ takes the form 
\begin{eqnarray}
\label{gammaope} 
&&\Gamma_{\rm OPE}(T,q)=\Gamma_0(T,q)+\Gamma_{\rm power}(T,q), 
\nonumber\\
&&\qquad \Gamma_{\rm power}(T,q)=\left(\frac{m}{2\pi T}\right)^{3/2} 
\left[ -\frac{1}{4}\omega^2T^2+\frac{q^2\omega^2}{24m}T^3 +
\left(\frac{19}{480}\omega^4-\frac{5q^4 \omega^2}{1536 m^2}\right)T^4+\cdots \right].
\end{eqnarray}
We display here only terms up to $O(T^4)$ in $\Gamma_{\rm power}$. 
For the practical extraction of the ground-state form factor we 
can retain any number of higher-order terms, which may be easily
generated from the exact expression (\ref{gamma}).

It should be emphasized that the coefficients of each power of
$T^n$ in the square brackets of (\ref{gammaope}) are
polynomials in $q^2$ of order $(n-2)$. Therefore, if the momentum
$q$ increases, one needs to include more and more power
corrections in order to achieve a certain accuracy of the
truncated OPE series for $\Gamma$. In QCD, this implies the
necessity to know and include condensates of higher and higher dimensions. 
Since only a few lowest-order condensates are known, in practice 
the applicability of three-point sum rules in QCD is restricted 
to the region of not too large $q^2$ and $T$.

\begin{figure}[!tb]
\begin{tabular}{ccc}
\includegraphics[width=5.8cm]{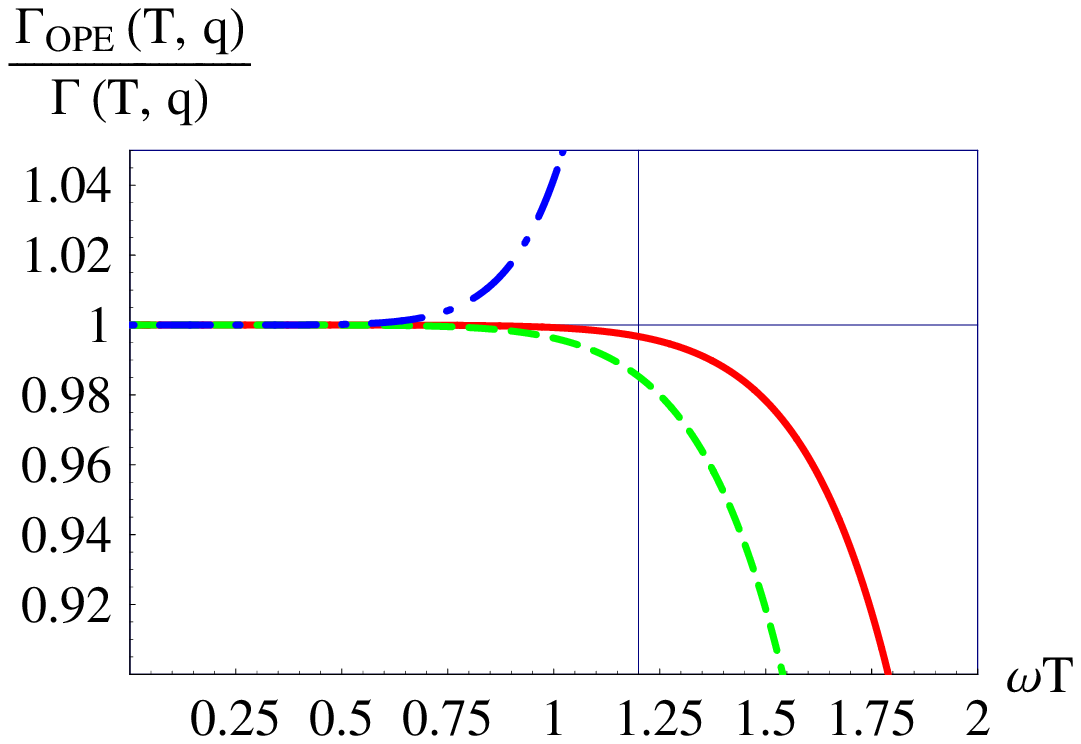}&
\includegraphics[width=5.8cm]{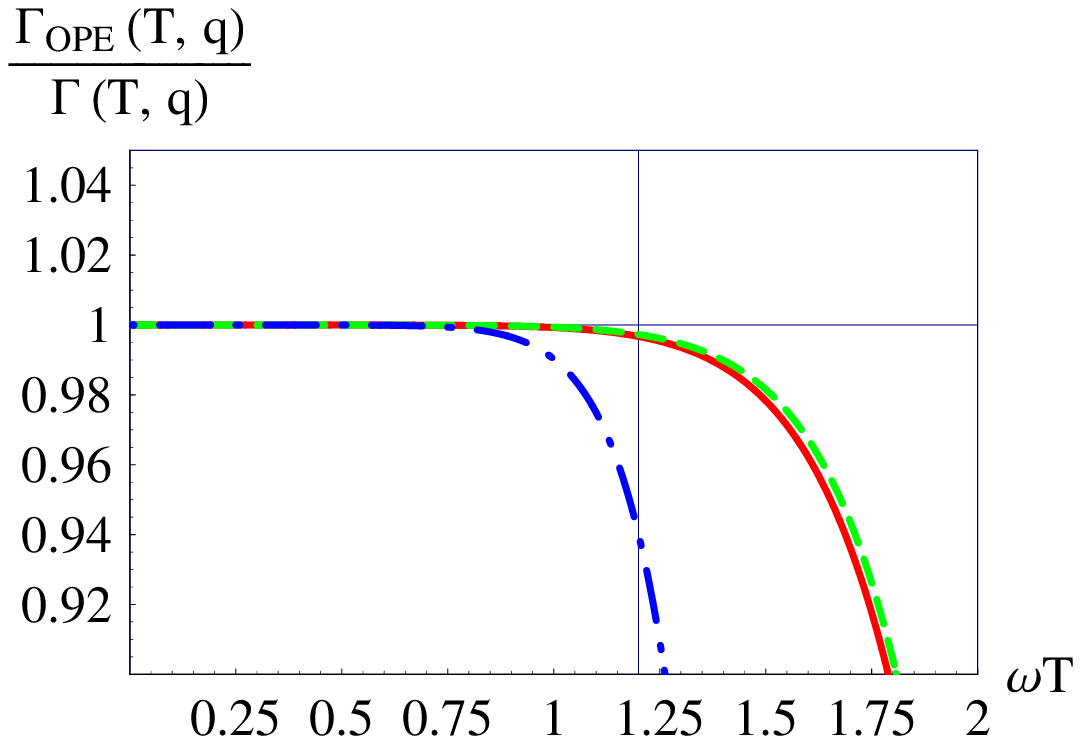}&
\includegraphics[width=5.8cm]{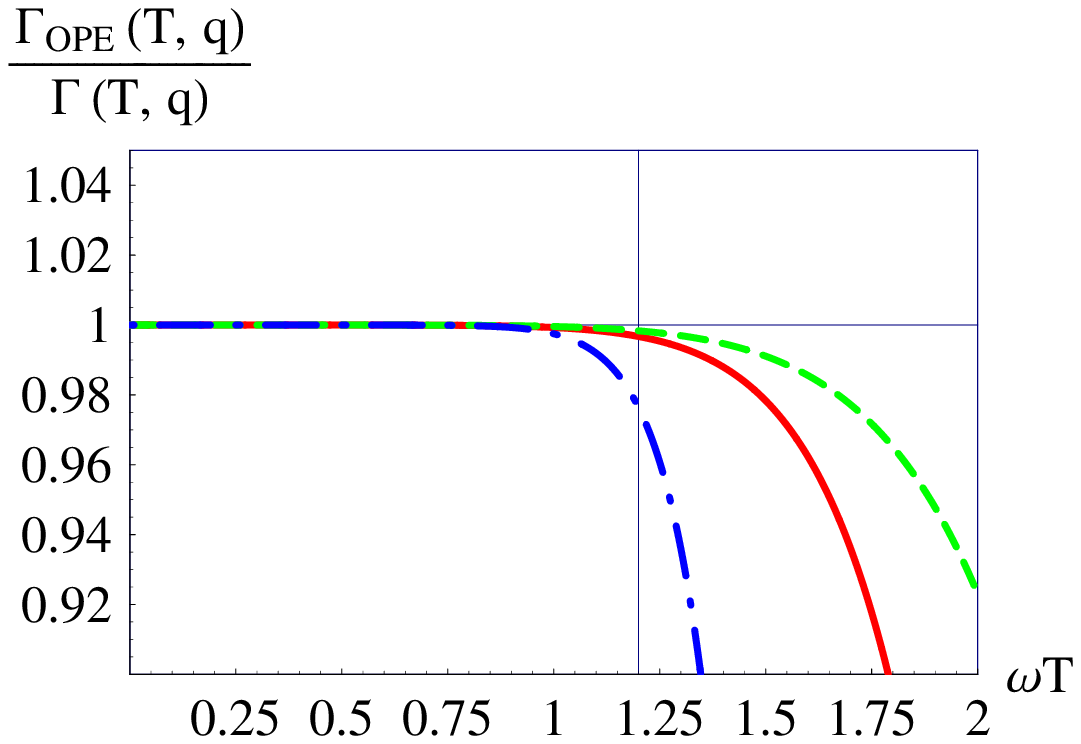}\\
(a)& (b) & (c)
\end{tabular}
\caption{\label{Plot:2} 
The accuracy of the truncated OPE for $\Gamma(T,q)$ depending on the number of retained 
power corrections, for different values of the momentum transfer $q$. 
The plots present $\Gamma_{\rm OPE}(T,q)/\Gamma(T,q)$ for the following cases:   
(a) the terms including $O(T^6)$ are retained in $\Gamma_{\rm OPE}$; 
(b) the terms including $O(T^8)$ are retained in $\Gamma_{\rm OPE}$; 
(c) the terms including $O(T^{10})$ are retained in $\Gamma_{\rm OPE}$.
Solid (red) line: $q=0$; dashed (green) line: $q=1.5\sqrt{m\omega}$; 
dash-dotted (blue) line: $q=3\sqrt{m\omega}$. 
The vertical line at $\omega T=1.2$ is the upper boundary of the Borel window. }
\end{figure}

Figure \ref{Plot:2} demonstrates the behaviour of the exact
correlator and the truncated OPE for different numbers of power corrections retained in the 
expansion and for several values of the momentum transfer $q$. 
Clearly, retaining a fixed number of power corrections and 
requiring the accuracy of the truncated OPE to be better than say 1\% for $\omega T\le 1.2$ 
restricts the region of the momentum transfer to ``not very large'' values. 
As is clear from Figs.~\ref{Plot:1} and \ref{Plot:2}, in the region of $\omega T$ where the truncated 
OPE provides a good description of the exact correlator, the contribution of the excited states is still rather 
large, i.e., one is still rather far from the plateau. Therefore, a direct determination of the
form factor from a truncated OPE is not possible. The procedures of the method of sum rules  
are aimed at {\it modeling} the contribution of higher states to the correlator and at 
obtaining in this way the ground-state form factor.


\section{\label{Sect:SR}Sum rules and exact effective continuum threshold}
The sum rules for the two-point correlator $\Pi$ and the three-point correlator $\Gamma$ are merely 
an expression of equality of these correlators calculated in the ``quark'' and in the hadron basis:
\begin{eqnarray}
\label{sr1}
R_{\rm g} e^{-{E_{\rm g}}T}              +\Pi_{\rm excited}(T)&=&\Pi_0(T) + \Pi_{\rm power}(T),\\
F_{\rm g}(q)R_{\rm g} e^{-{E_{\rm g}}T}+\Gamma_{\rm excited}(T,q)&=&\Gamma_0(T,q) + \Gamma_{\rm power}(T,q). 
\end{eqnarray}
Making use of the standard quark-hadron duality assumption --- that the contribution of the
ground state is dual to the low-energy region --- we obtain the following duality relations:\footnote{
The standard duality assumption may be formulated as follows: the contribution of higher hadron 
states is dual to the high-energy region of the free-quark diagrams. 
Therefore, excited states do not receive any contribution from power corrections. 
One may, of course, argue whether this assumption is physically relevant: it would be perhaps more natural 
to relate to the excited states also high-energy contribution of diagrams containing the confinement 
interaction, like those corresponding to $\Gamma_2$. 
This would, however, require a detailed 
knowledge of the confining potential.}
\begin{eqnarray}
\label{srpi}
R_{\rm g}e^{-{E_{\rm g}}T}
&=&
\int\limits_{0}^{z^\Pi_{\rm eff}(T)}dz\,e^{-z T}\rho_0(z)+ \Pi_{\rm power}(T)\equiv \Pi_{\rm dual}(T,z^\Pi_{\rm eff}(T)),
\\
\label{srgamma}
F_{\rm g}(q)R_{\rm g}e^{-{E_{\rm g}}T}
&=&
\int\limits_{0}^{z_{\rm eff}(T,q)}dz_1 \int\limits_{0}^{z_{\rm eff}(T,q)}dz_2\,e^{-\frac12 z_1T}e^{-\frac12 z_2T}
\Delta_0(z_1,z_2,q)+ \Gamma_{\rm power}(T,q)\equiv \Gamma_{\rm dual}(T,q,z_{\rm eff}(T,q)).\nonumber\\
\end{eqnarray}
We call the r.h.s. of (\ref{srpi}) and (\ref{srgamma}) the dual 
correlators $\Pi_{\rm dual}(T,z^\Pi_{\rm eff}(T))$ and $\Gamma_{\rm dual}(T,q,z_{\rm eff}(T,q))$, respectively. 
Let us notice that the dual correlators have both 
an explicit dependence on $T$ and an implicit dependence on $T$ via $z_{\rm eff}$. 

As the consequence of (\ref{pigamma}), $\Pi_{\rm power}(T)=\Gamma_{\rm power}(T,q=0)$. It is natural to set  
\begin{eqnarray}
z^\Pi_{\rm eff}(T)=z_{\rm eff}(T,q=0). 
\end{eqnarray} 
Then, due to the Ward identity (\ref{wi}), the relations (\ref{srpi}) and 
(\ref{srgamma}) yield the correct normalization of the form factor 
$F_{\rm g}(q=0)=1$, as required by current conservation. 

\begin{figure}[!htb]
\begin{tabular}{cc}
\includegraphics[width=8.2cm]{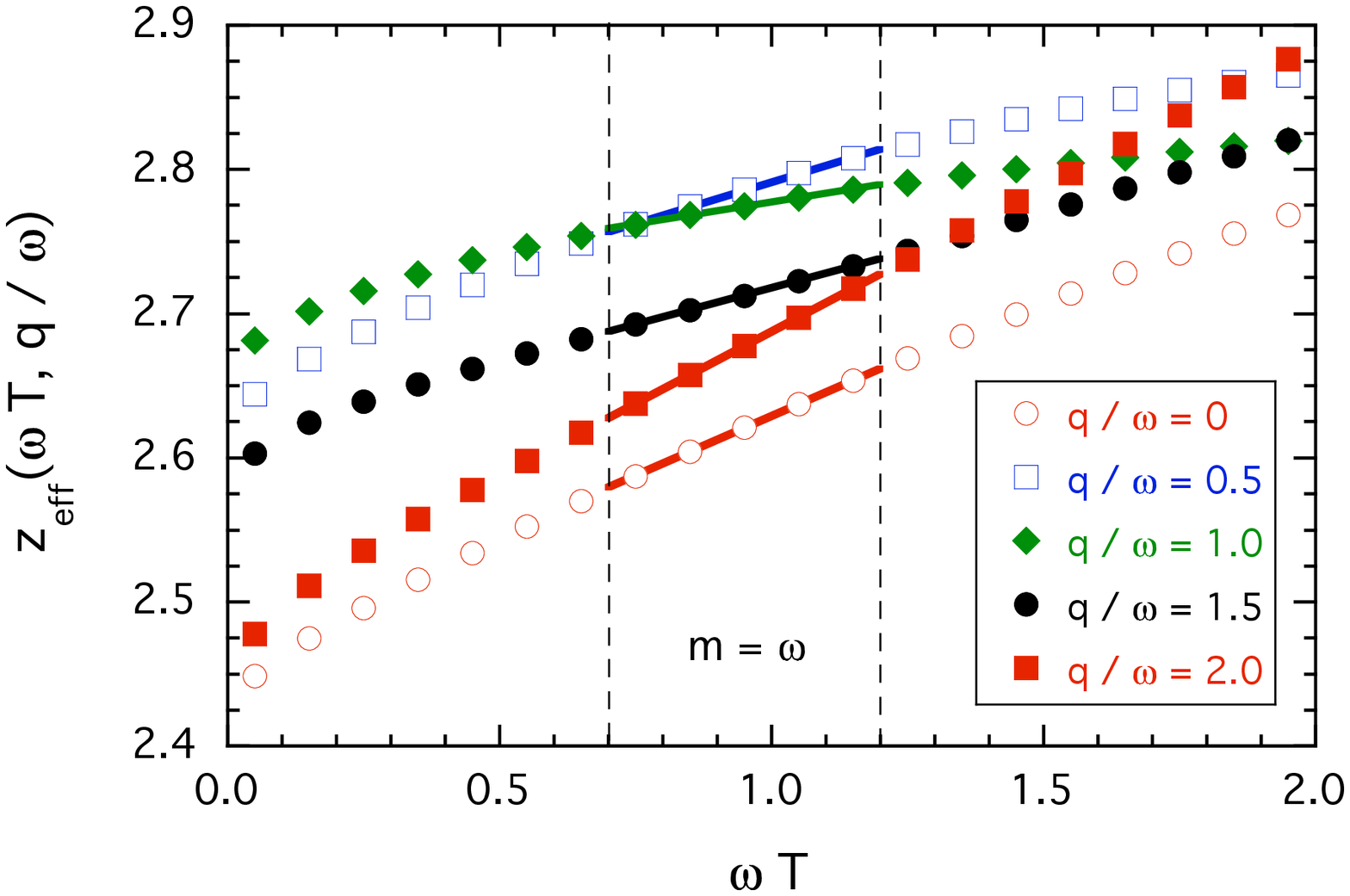}&
\includegraphics[width=8.2cm]{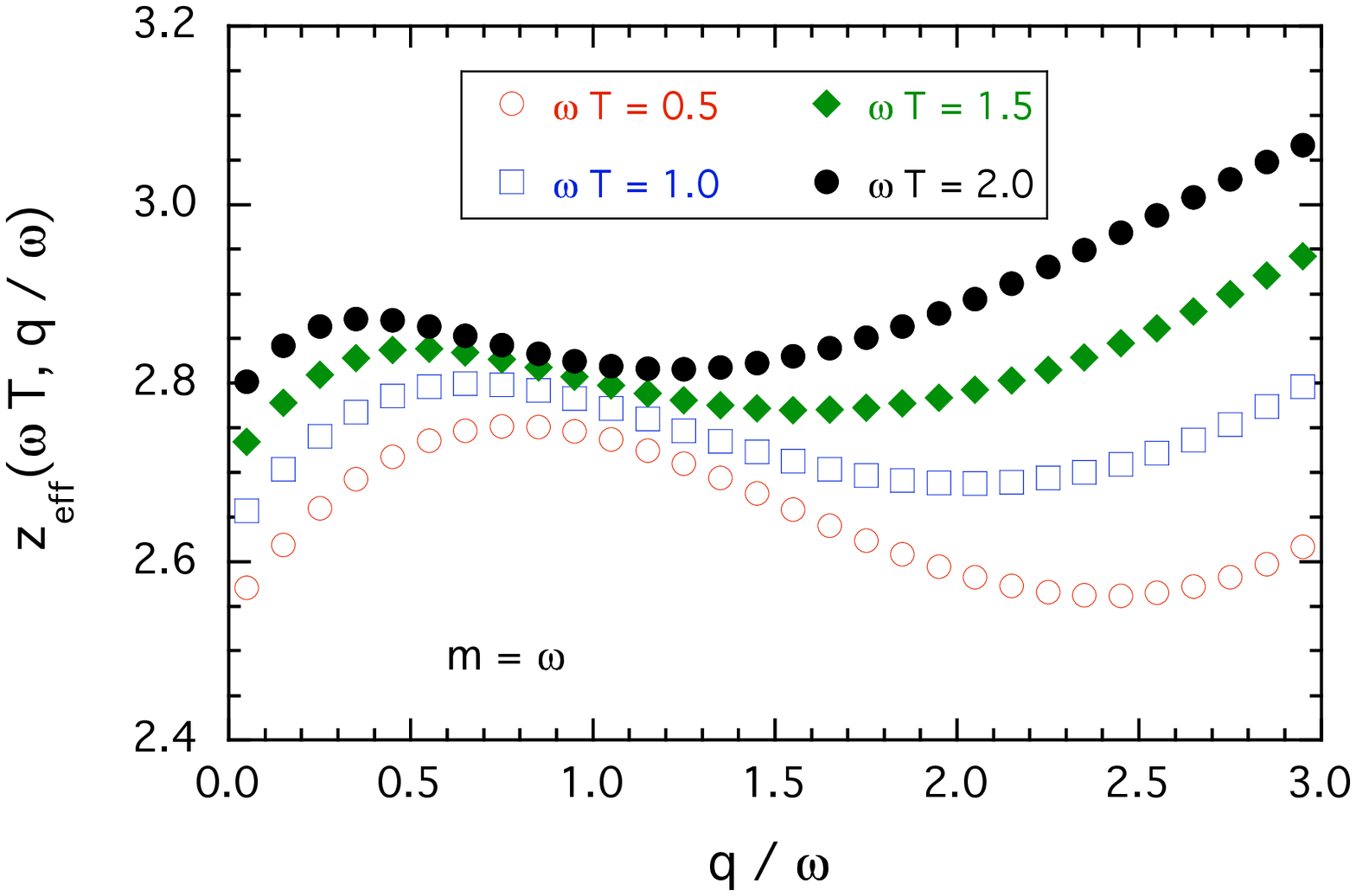}\\
(a)& (b) 
\end{tabular}
\caption{\label{Plot:3}
The effective continuum threshold $z_{\rm eff}(T, q)$ for the 3-point function, obtained 
by solving numerically Eq.~(\ref{srgamma}) using the exact bound-state parameters 
$R_{\rm g}$ and $F_{\rm g}(q)$ as well as the exact power expansion $\Gamma_{\rm power}(T, q)$, 
versus the Euclidean time $T$ at fixed values of the momentum transfer $q$ (a) 
and versus $q$ at fixed values of $T$ (b).
In (a) the vertical dashed lines identify the {\em fiducial range} in $T$ (see 
text), while the solid lines are linear fits of $z_{\rm eff}(T, q)$ in the 
fiducial range.}
\end{figure}

We treat the expressions (\ref{srpi}) and (\ref{srgamma}) as the definitions of the exact $T$- and
$q$-dependent effective continuum thresholds $z_{\rm eff}(T, q)$ and $z_{\rm eff}^{\Pi}(T)$ corresponding to 
the true ground-state parameters on the l.h.s. The full $T$- and $q$-dependences of 
$z_{\rm eff}(T, q)$ can be obtained by solving Eqs.~(\ref{srpi}) and (\ref{srgamma}) for the known 
exact bound-state parameters $R_{\rm g}$ and $F_{\rm g}(q)$ and the exact power 
expansions $\Pi_{\rm power}(T)$ and $\Gamma_{\rm power}(T, q)$. In the HO model this can be easily 
done numerically. 
Without loss of generality we set $m = \omega$ and show the corresponding results in Fig.~\ref{Plot:3}.
Obviously, the effective continuum threshold does depend sizeably on both $T$ and $q$.
Now, a good or a bad extraction of the ground-state parameters depends on our capacity to find 
a reasonable approximation to the exact effective continuum threshold.

Before closing this Section, we would like to make a comment on the $T$-dependence of 
the effective continuum threshold and the analytic properites of the dual correlator. 

Recall that a correlator in perturbation theory satisfies the standard dispersion representation 
in $E$ with the cuts given by the quark diagrams. In the language of hadronic states, 
this correlator contains contributions of an infinite number of such states. 

The dual correlator is a hand-made object and its analytic properties are very much 
different: the dual correlator should reproduce the contribution of a single ground state only, 
so it should have a single pole in $E$ and no cuts. Equivalently, the Borel-transformed 
dual correlator should contain only single exponential $\exp(-E_{\rm g}T)$. 
Obviously, this may happen only if the effective threshold has a complicated dependence 
on $T$, since the spectral densities of the dual correlator and the perturbative 
correlator coincide. 

Therefore, the dependence of the effective threshold on $T$ is its inherent property required 
by the analytic properties of the dual correlator. 

\section{\label{Sect:Numerics}Numerical analysis}
Let us now consider a restricted problem: assume that we know exactly the energy of the ground state, 
$E_{\rm g}$. Our goal is to determine the decay constant and the form factor of this state from 
the sum rules (\ref{srpi}) 
and (\ref{srgamma}). 
This problem is similar to the realistic case of the $B$-meson observables: 
in this case one knows the mass of the state and attempts to calculate its characteristics. 

To obtain theoretical estimates for ground-state parameters, we must perform the following steps: 

First, according to \cite{svz} we should determine the Borel window (or the fiducial range), where 
according to the sum-rule philosophy the method may be applied to the extraction of the 
ground-state parameters: 
(i) the lower boundary of the $T$-window is found from the requirement that the 
ground state gives a sizable (we require more than 50\%) contribution to the 
correlator; and
(ii) the upper boundary of the $T$-window is obtained from the requirement that the 
truncated OPE gives a good (we require better than 1\%) approximation to 
the exact correlator. 

Second, we must choose a criterion to fix our approximation for the effective continuum threshold 
$z_{\rm eff}(T, q)$ and obtain in this way our sum-rule estimates for hadron observables. 

Third, we should define the way to obtain the band of values which contains the 
actual value with 100\% probability and a flat probability distribution,  
in other words, define the way to obtain systematic errors of the hadron parameter.  
Obviously, this is a highly nontrivial point. 

The standard procedure adopted by sum-rule practitioners is to assume a $T$-independent 
effective continuum threshold. 
The quantity may be then either chosen as a $q$-independent constant or 
adjusted for any value of $q$, separately. 

The variation of the extracted form factor in the window is conventionally treated as the systematic 
error of the ground-state parameter. It should be recalled that in QCD this is a matter of belief: 
it is not possible to prove that the interval obtained in this way indeed contains the true value. 

In a series of recent publications we have demonstrated that in many cases these standard criteria fail: 
the actual parameters of the ground state turn out to lie beyond the range predicted by the sum-rule 
procedures. 
In \cite{lms_2ptsr} we have shown this for the extraction of the ground-state decay constant from the 
two-point function; 
in \cite{lms_3ptsr,m_lcsr} we have given examples of the form factor at specific values 
of momentum transfers, 
for which the standard criteria (stability in the window, deviation of the energy of the cut correlator 
from the ground-state energy in the window) suggest an accurate extraction of the form factor, 
whereas in 
practice the actual error turns out to be much larger. Now, we are going to investigate whether 
improvements 
may be obtained by releasing the requirement of the $T$-independent effective continuum threshold and by 
imposing other criteria than Borel stability for obtaining the error estimates. 

In this work we discuss three different Ans\"atze for the effective continuum threshold:
\begin{eqnarray}
\label{constant}
z_{\rm eff}(T, q) & \approx & z_0^C(q), \\
\label{linear}
z_{\rm eff}(T, q) & \approx & z_0^L(q) + z_1^L(q) ~ \omega T, \\
\label{quadratic}
z_{\rm eff}(T, q) & \approx & z_0^Q(q) + z_1^Q(q) ~ \omega T + z_2^Q(q) ~ \omega^2 T^2.
\end{eqnarray}
At each value of $q$ we fix the unknown parameters on the r.h.s of 
Eqs.~(\ref{constant})--(\ref{quadratic}) in the following way: 
we define the dual energy, $E_{\rm dual}(T, q)$, as
\begin{eqnarray}
\label{Edual}
E_{\rm dual}(T, q) =  - \frac{d}{d T} ~ \mbox{log} ~ \Gamma_{\rm dual}(T, q, z_{\rm eff}(T, q)),
\end{eqnarray}
where $\Gamma_{\rm dual}(T, q, z_{\rm eff})$ is the r.h.s.~of Eq.~(\ref{srgamma}) calculated 
using the approximations (\ref{constant})--(\ref{quadratic}) for $z_{\rm eff}(T, q)$.
Let us emphasize that the implicit dependence on $T$ via $z_{\rm eff}$ is crucial for the 
calculation of the energy: if the exact effective continuum threshold has a sizeable 
dependence on $T$, neglecting this dependence leads to a completely wrong energy, 
which we later tune to the exact known value. This is, in fact, the main source of error 
in the sum-rule predictions for hadron observables. 

Next, we calculate $E_{\rm dual}(T, q)$ at several values of $T = T_i$ ($i = 1,\dots, N$) 
chosen uniformly in the fiducial range. Finally, we minimize the squared 
difference between the dual energy $E_{\rm dual}(T, q)$ and the exact value $E_{\rm g}$:
\begin{eqnarray}
\label{chisq}
\chi^2 \equiv \frac{1}{N} \sum_{i = 1}^{N} \left[ E_{\rm dual}(T_i, q) - E_{\rm g}\right]^2.
\end{eqnarray}

\subsection{Decay constant}
Let us start with $R_{\rm g}$, the square of the decay constant. 
The standard procedures for this case were discussed 
in detail in \cite{lms_2ptsr}. The main results of \cite{lms_2ptsr} may be summarized as follows: 
in spite of the fact that the numerical value of the extracted 
decay constant turns out to be not far from the exact value (see also the discussion in \cite{bakulev}), 
the standard procedures fail to produce realistic error estimates: 
namely, {\it the known true value lies outside the range obtained from  
the standard sum-rule analysis.} 

\begin{figure}[!htb]
\begin{tabular}{ccc}
\includegraphics[width=5.5cm]{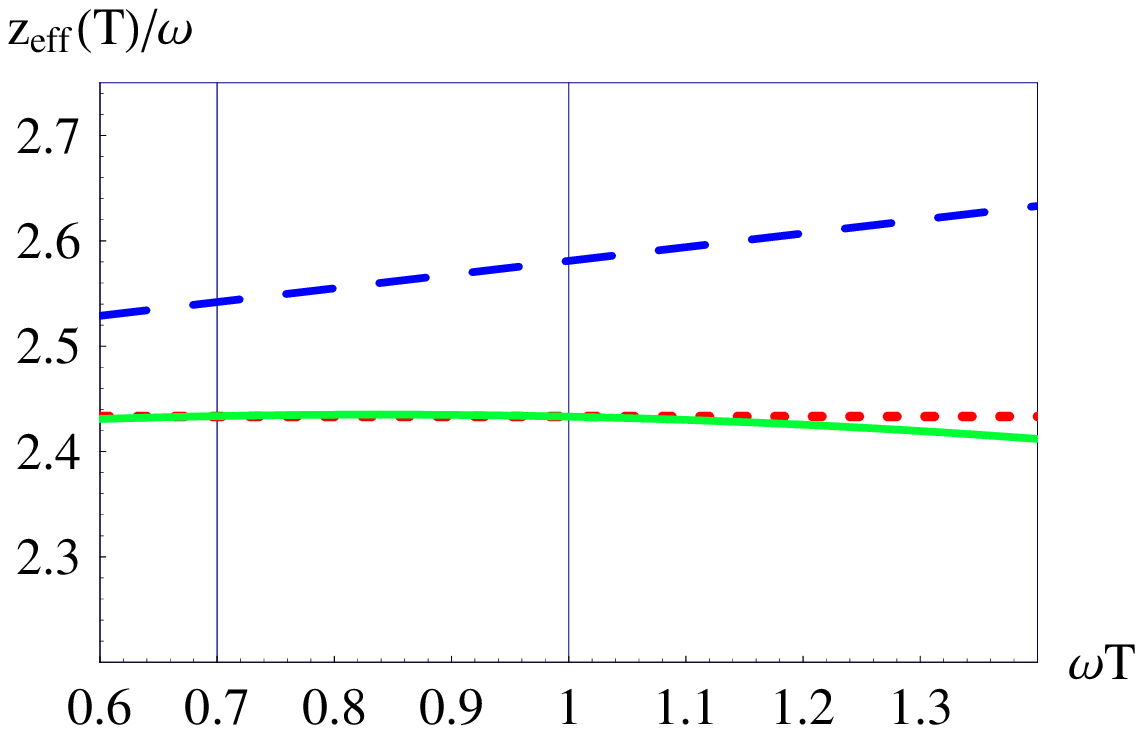}&
\includegraphics[width=5.5cm]{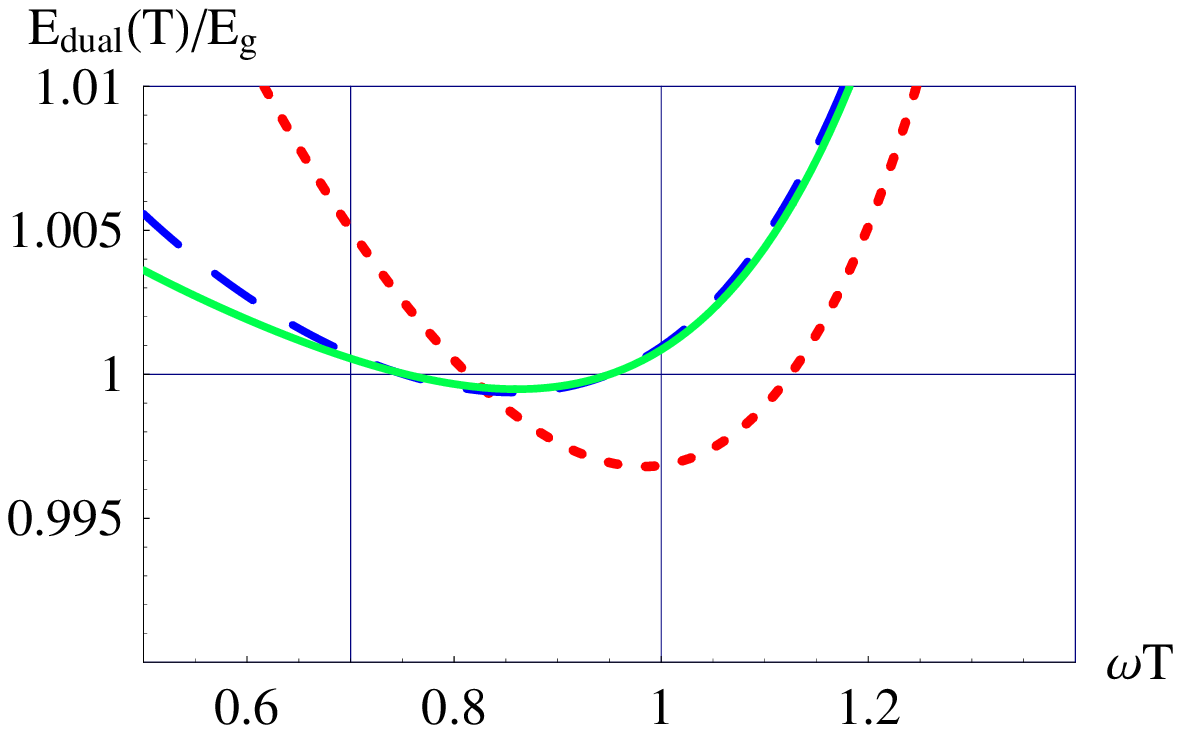}&
\includegraphics[width=5.5cm]{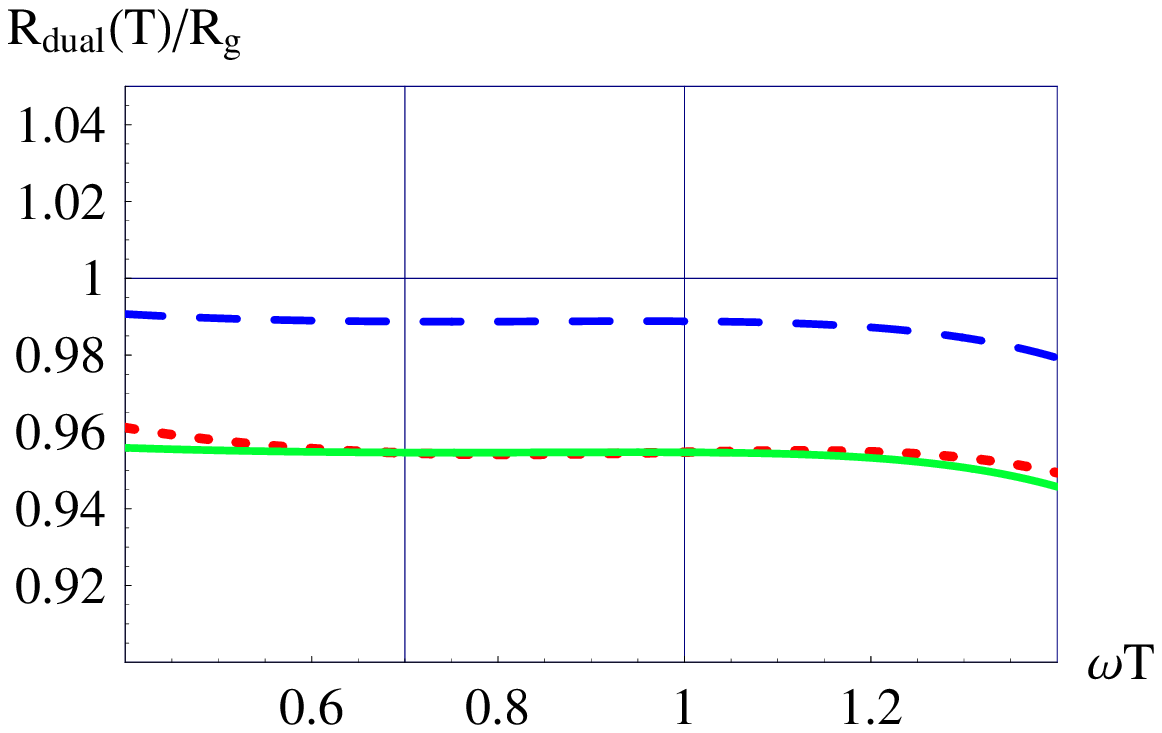}\\
\includegraphics[width=5.5cm]{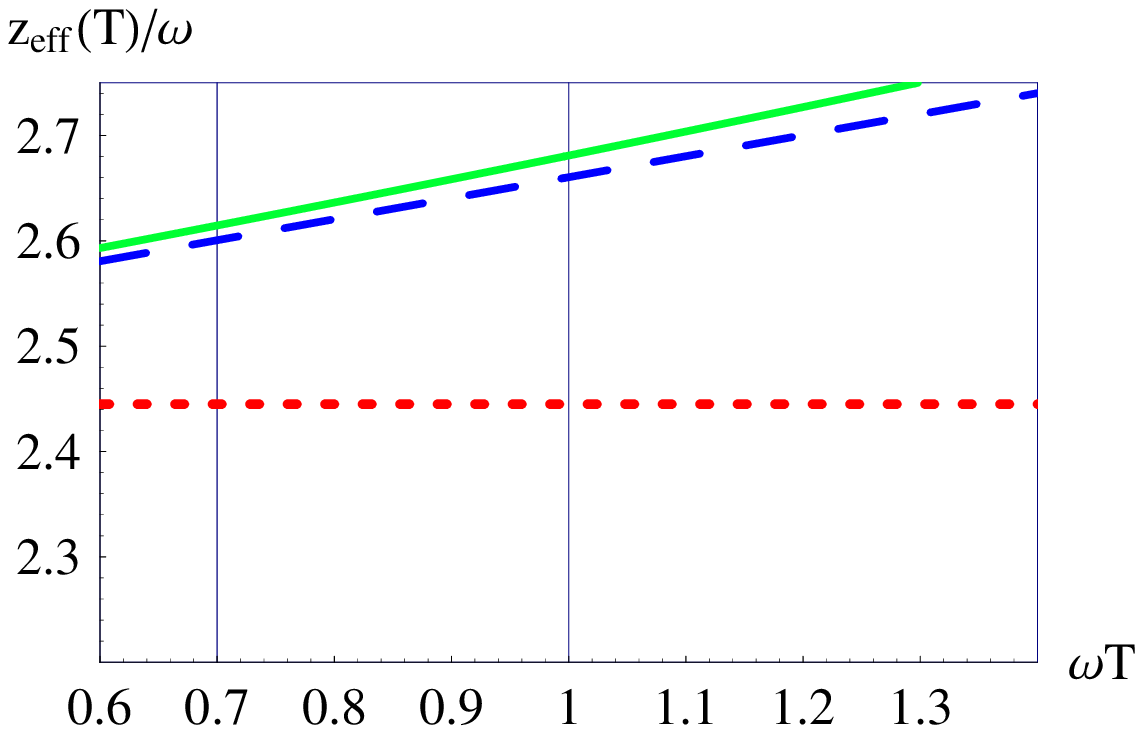}&
\includegraphics[width=5.5cm]{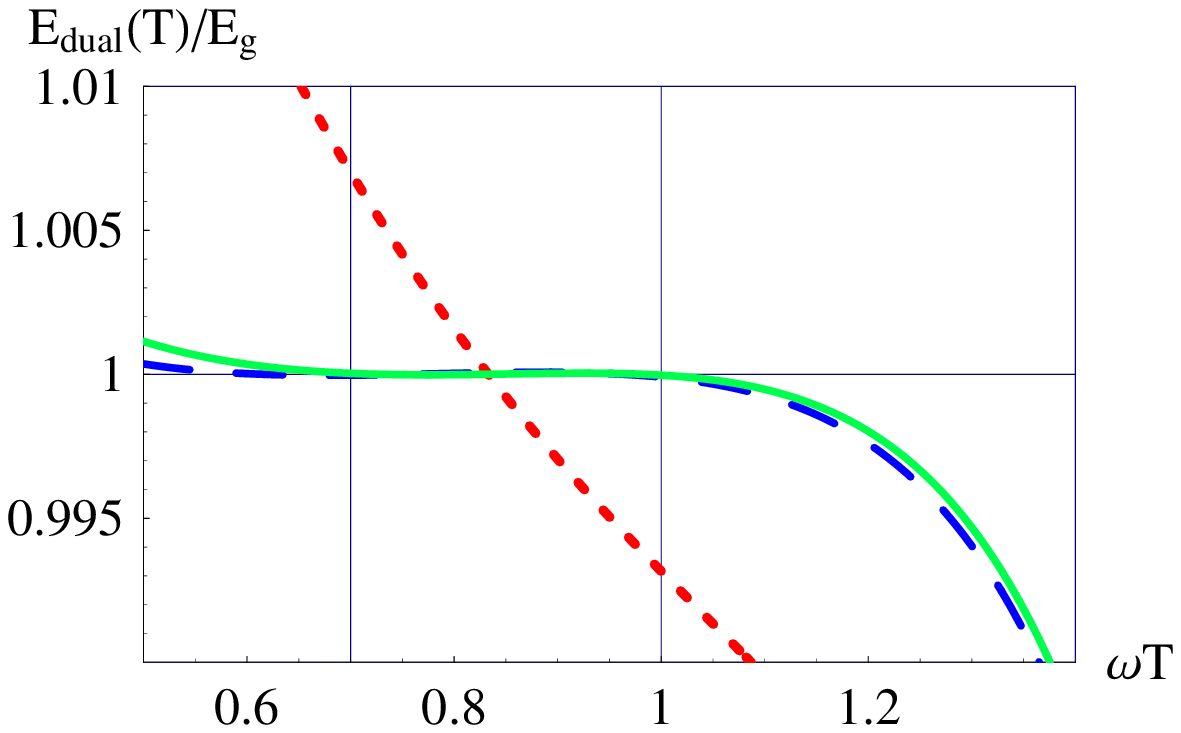}&
\includegraphics[width=5.5cm]{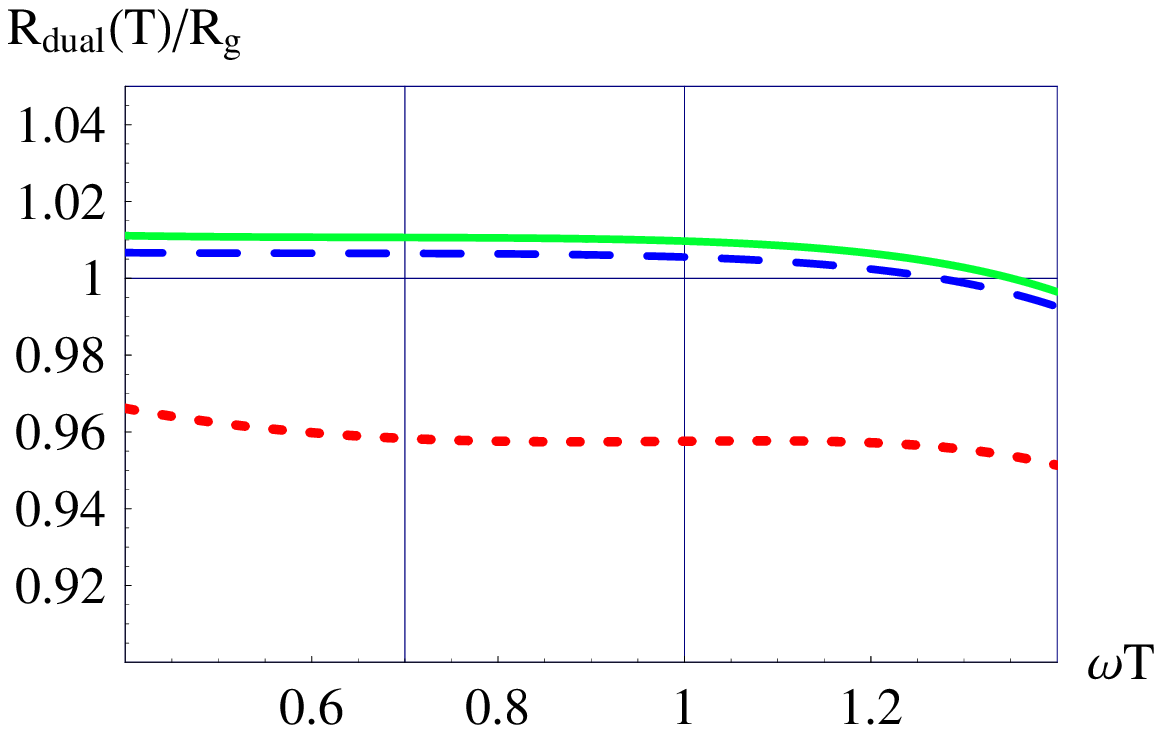}\\
\includegraphics[width=5.5cm]{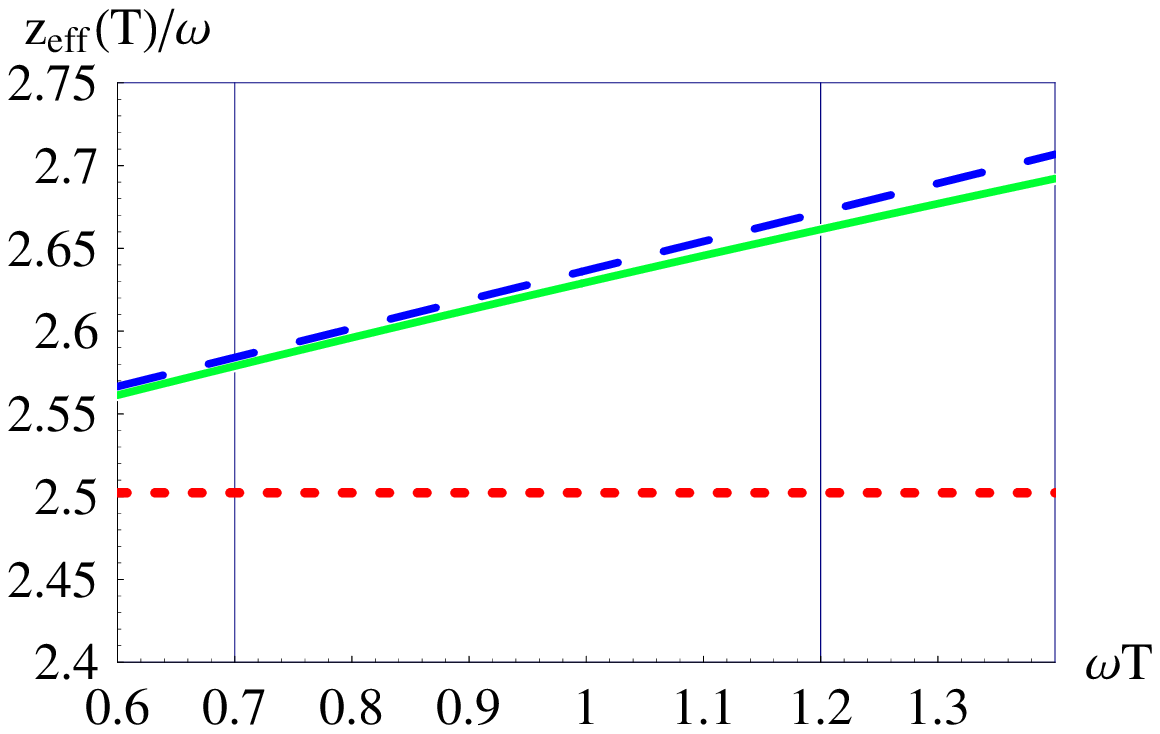}&
\includegraphics[width=5.5cm]{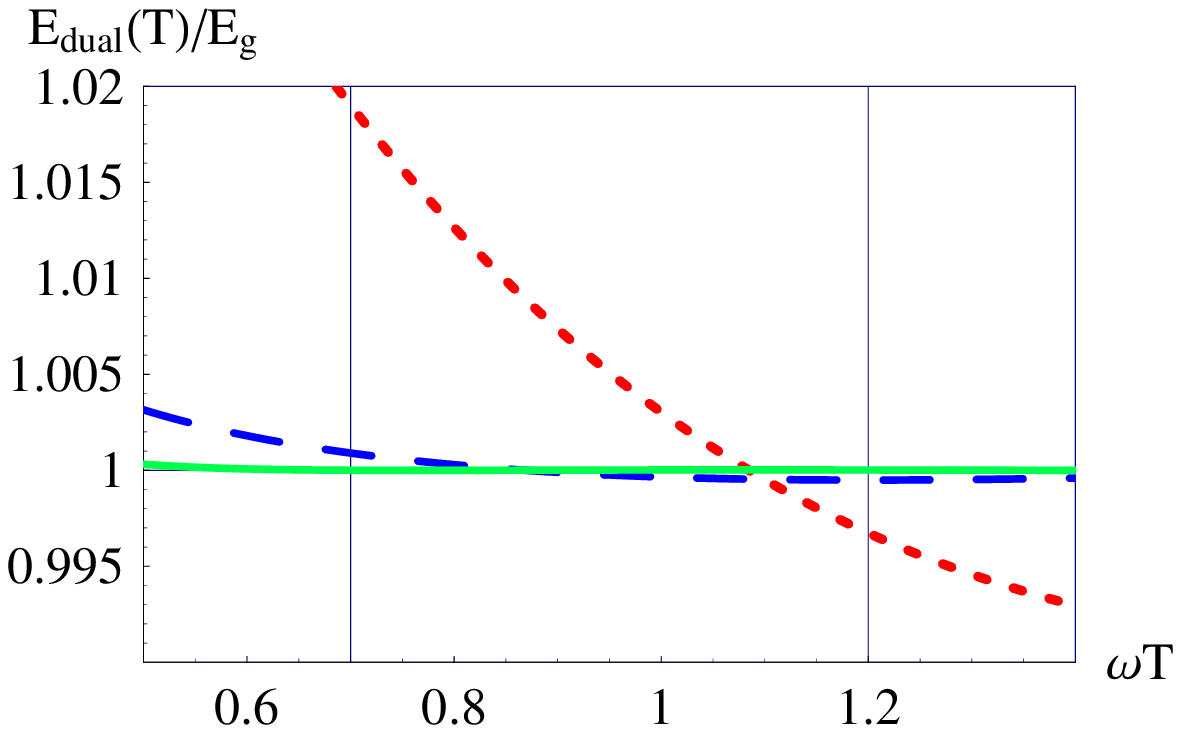}&
\includegraphics[width=5.5cm]{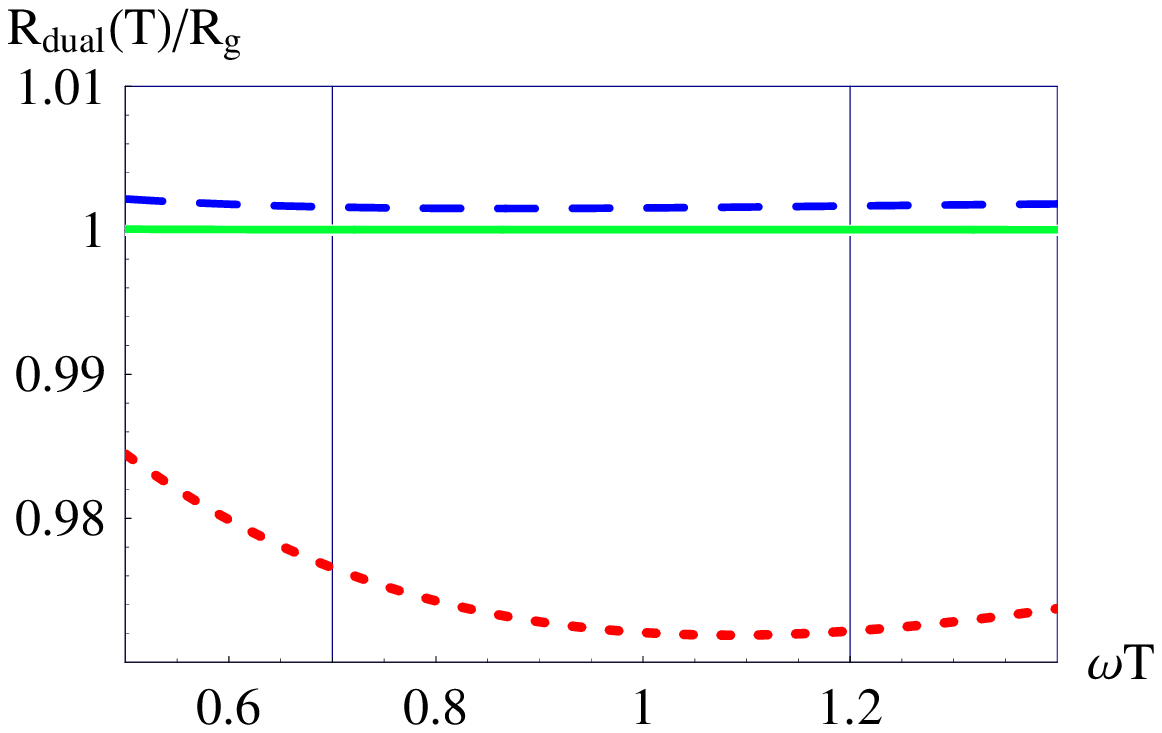}\\
\end{tabular}
\caption{\label{Plot:4}
The extraction of $R_{\rm g}$, the square of the decay constant of the ground state, 
making use of $N$ power corrections in $\Pi_{\rm OPE}(T)$: $N=3$ (first row), $N=4$ (second row), and 
$N=\infty$ (third row). First column: the constant, linear, and quadratic approximations for 
the exact effective threshold $z^\Pi_{\rm eff}(T)$ obtained 
by minimizing (\ref{chisq}).
Second column: the corresponding $E_{\rm dual}(T)/E_{\rm g}$.  
Third column: the corresponding $R_{\rm dual}(T)/R_{\rm g}$. 
Red (dotted): constant $z^C_{\rm eff}$; 
blue (dashed): linear  $z^L_{\rm eff}(T)$; 
green (solid): quadratic $z^Q_{\rm eff}(T)$. The vertical lines indicate the boundaries of the window. }
\end{figure}

Fig.~\ref{Plot:4} shows the results obtained by allowing for a $T$-dependent 
threshold and tuning its parameters according to (\ref{chisq}): one can see three approximations for 
$z^\Pi_{\rm eff}(T)$ obtained by minimizing the function (\ref{chisq}) for 
\begin{eqnarray}
\label{Edualpi}
E_{\rm dual}(T) =  - \frac{d}{d T} ~ \mbox{log} ~ \Pi_{\rm dual}(T, z^{\rm \Pi}_{\rm eff}(T)),
\end{eqnarray}
$\Pi_{\rm dual}(T, z^{\rm \Pi}_{\rm eff})$ being the r.h.s. of (\ref{srpi}), and 
the corresponding $R_{\rm dual}(T)/R_{\rm g}$. The extraction of $R$ is presented for three 
cases where a different number of power corrections --- 
three, four, and infinite (i.e., exact power corrections) --- in $\Pi_{\rm OPE}$ 
are taken into account. 

The case of the exact power corrections is presented as an illustration: 
we extract the ground-state parameters from the correlator by applying the sum-rule machinery in the window 
$0.7\le \omega T\le 1.2$ 
(although in this case, of course, the parameters may be extracted directly 
from the large-$T$ behaviour of this correlator). 

In all cases one can see an obvious improvement when the linear Ansatz $z^L_{\rm eff}(T)$ instead of the constant 
$z^C_{\rm eff}$ is used. Going beyond the linear approximation and allowing for the quadratic function 
$z^Q_{\rm eff}(T)$, however, does not lead to further improvement in the realistic cases of three and four 
power corrections in $\Pi_{\rm OPE}$. This effect is just the reflection of the fact that the ground-state parameter 
may be extracted from the correlator only with a limited accuracy, even if the ground-state mass is known precisely 
(see also the detailed discussion in \cite{lms_2ptsr}). 
The same situation occurs for the form factor in Section \ref{Sect:6B}. 

\subsection{Form factor\label{Sect:6B}}
The standard sum-rule analysis of the ground-state form factor was presented in \cite{lms_3ptsr}.
We have given there an explicit example of the sum-rule extraction of the form factor at a specific 
value of the momentum transfer where the method clearly fails in the following sense: 
the true value of the form factor turned out to lie well beyond the interval obtained by the standard 
sum-rule procedures. Moreover, the actual errors in the case 
of the form factor were shown to be much larger than those for the case of the decay constant. 

We now apply a $T$-dependent Ansatz for the effective continuum threshold 
to extract the form factor. Recall that because of current conservation, the form factor $F(q)$ 
should obey the absolute normalization $F(q = 0) = 1$. We therefore require 
$z^{\Pi}_{\rm eff}(T)=z_{\rm eff}(T,q=0)$: 
then at $q=0$ the r.h.s. of Eq.~(\ref{srgamma}) reproduces the r.h.s. of Eq.~(\ref{srpi}) and 
the form factor is automatically properly normalized. Therefore, the form factor extracted 
from the sum rule at small momentum transfers is anyway not far from its true value. Of real interest 
is the extraction of the form factor at the intermediate momentum transfers $q/\omega=1\div 2$.

In the HO model we know all power corrections exactly, therefore, we have no limitation on the upper 
boundary of the window. However, to be in line with the realistic situation where only a limited 
number of power corrections is available, we define the window to be $0.7 \lesssim \omega T \lesssim 1.2$ 
(see Fig.~\ref{Plot:2}).

\begin{figure}[t]
\begin{tabular}{ccc}
\includegraphics[width=5.5cm]{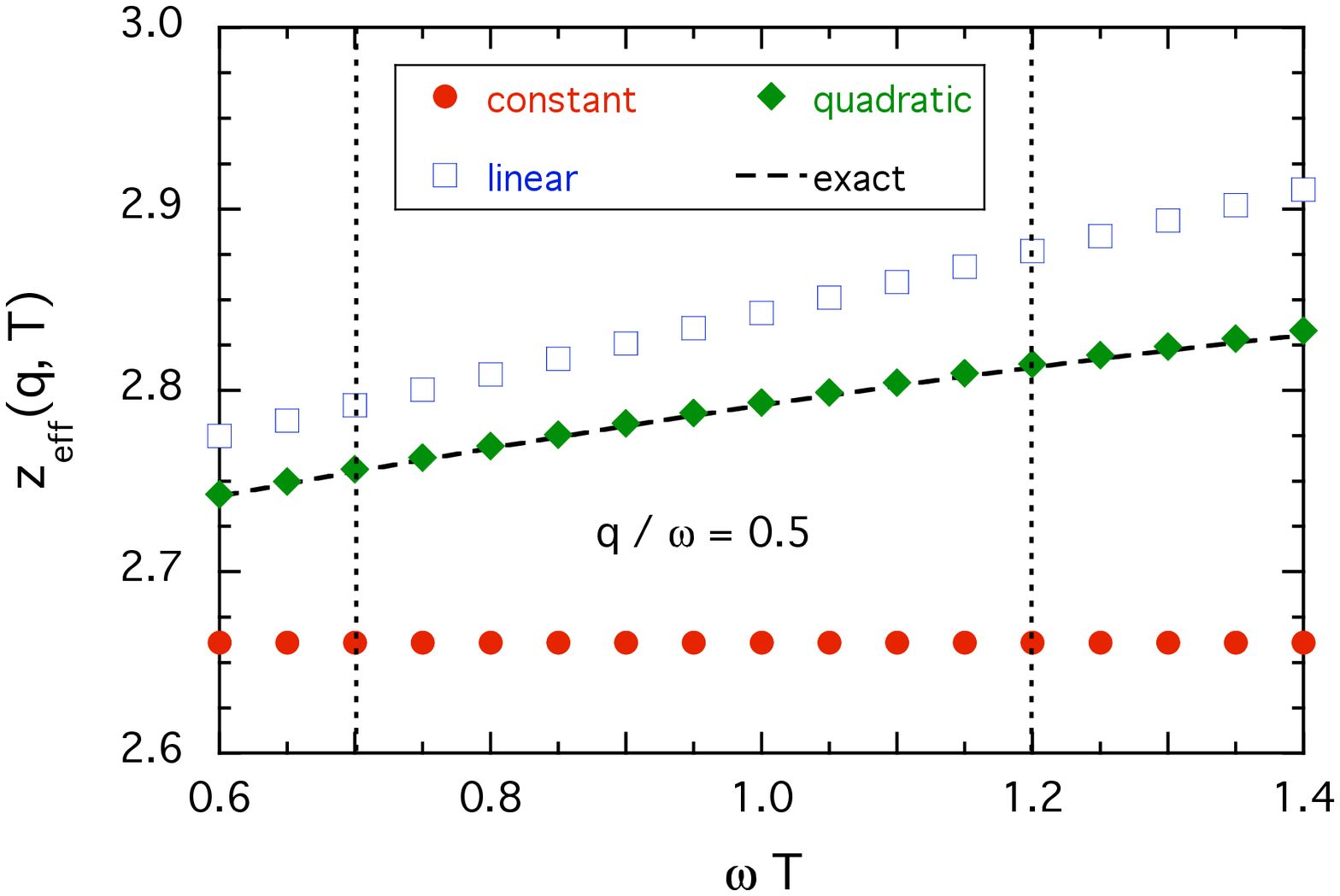}&
\includegraphics[width=5.5cm]{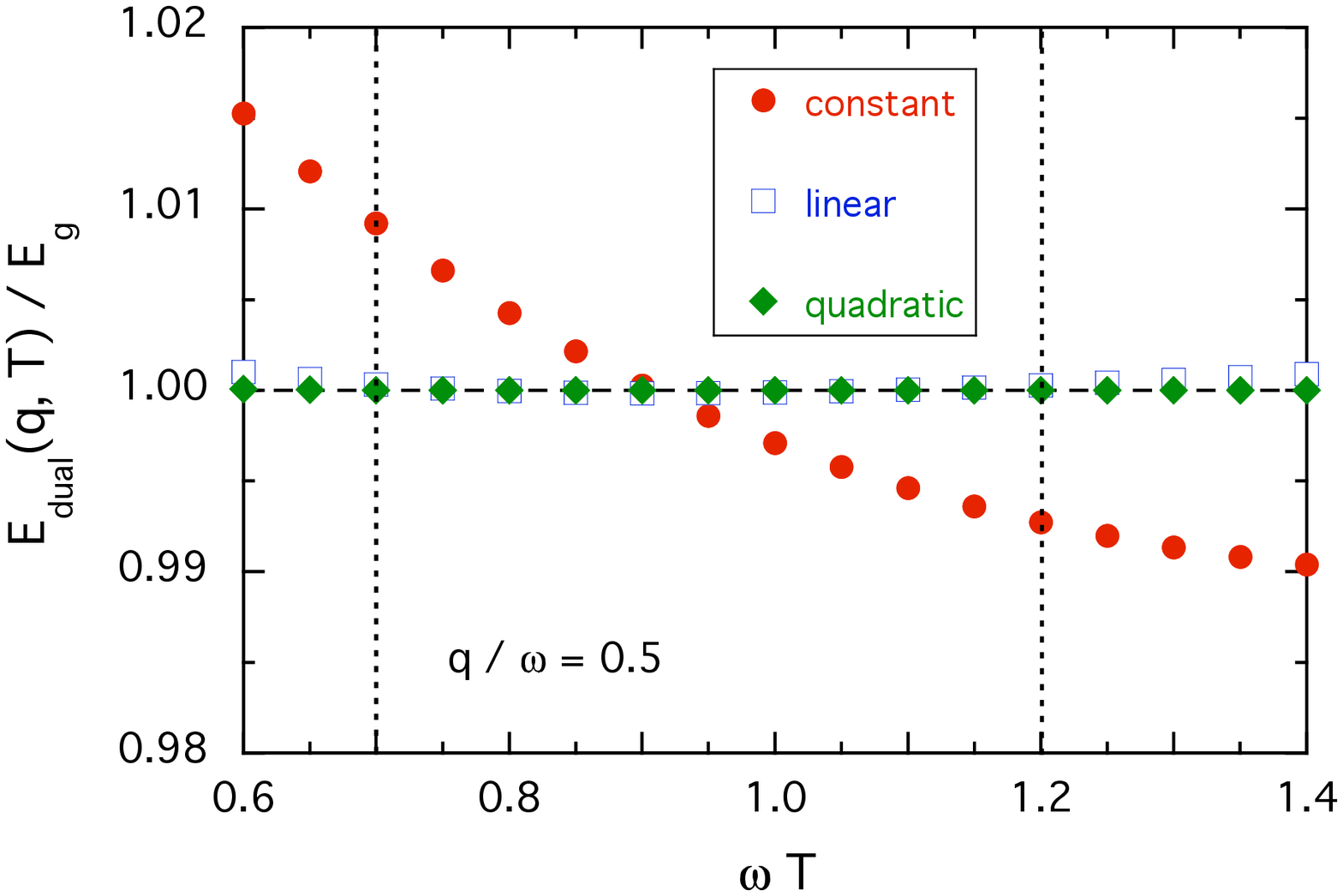}&
\includegraphics[width=5.5cm]{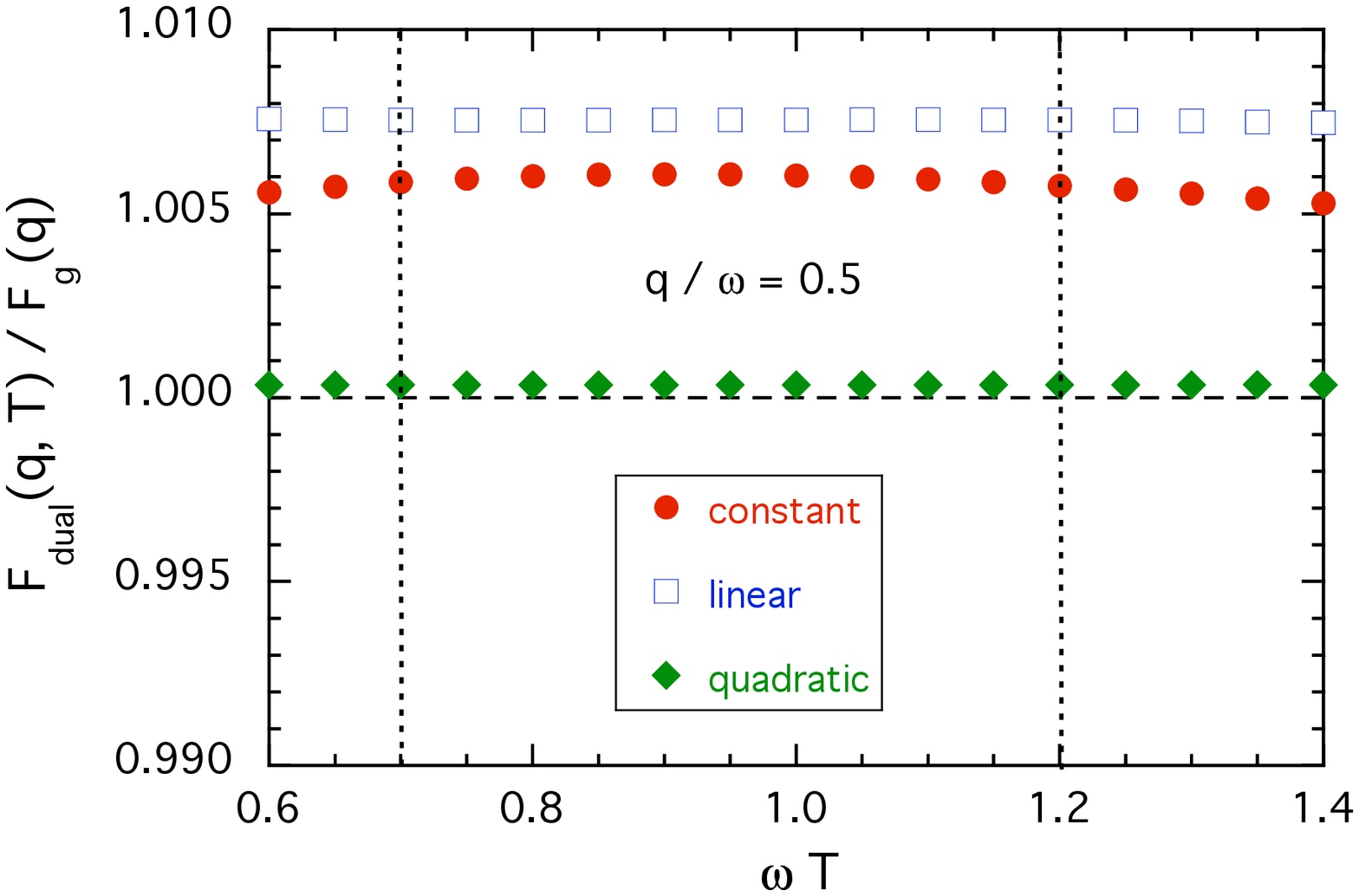}\\
\includegraphics[width=5.5cm]{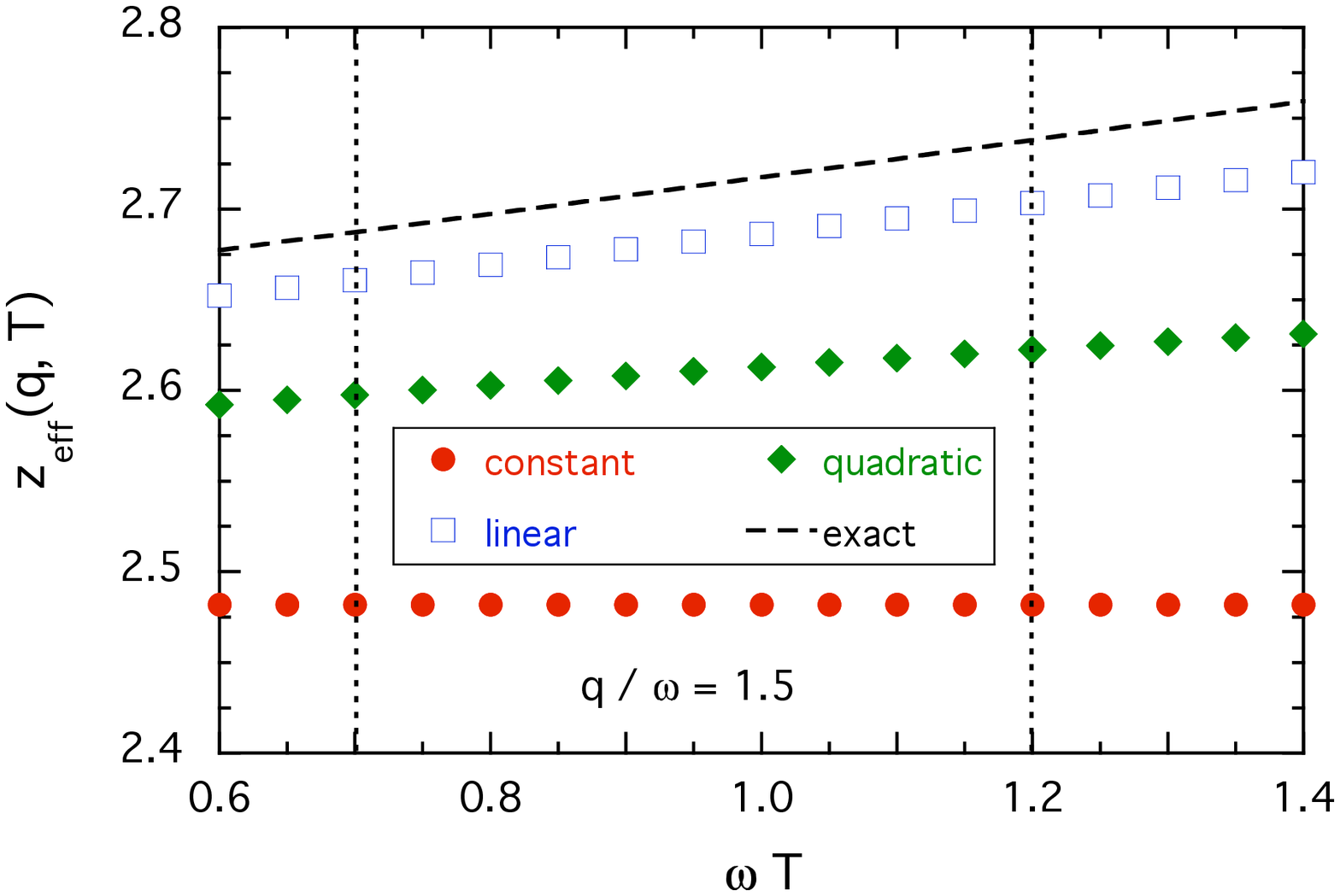}&
\includegraphics[width=5.5cm]{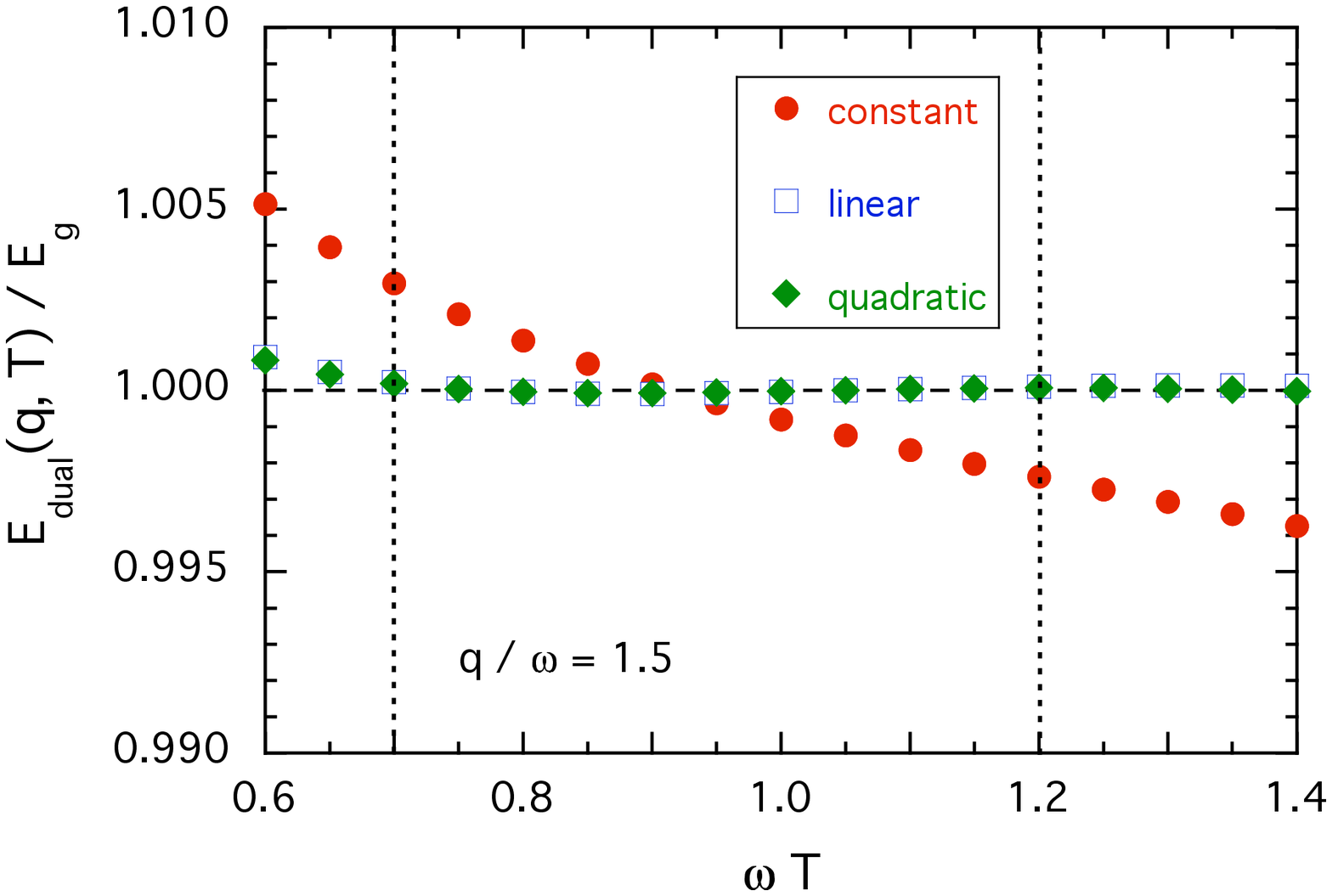}&
\includegraphics[width=5.5cm]{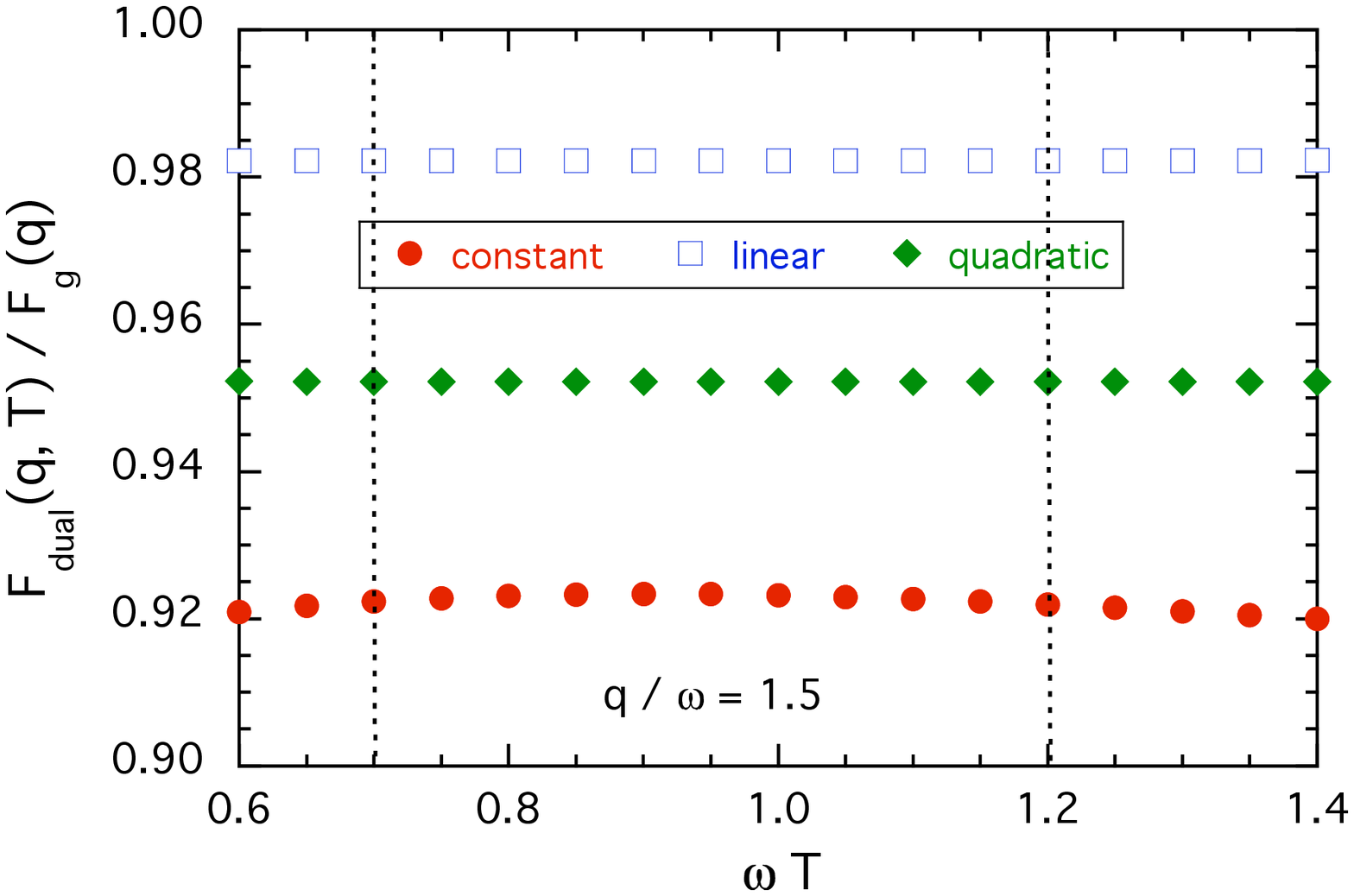}\\
\includegraphics[width=5.5cm]{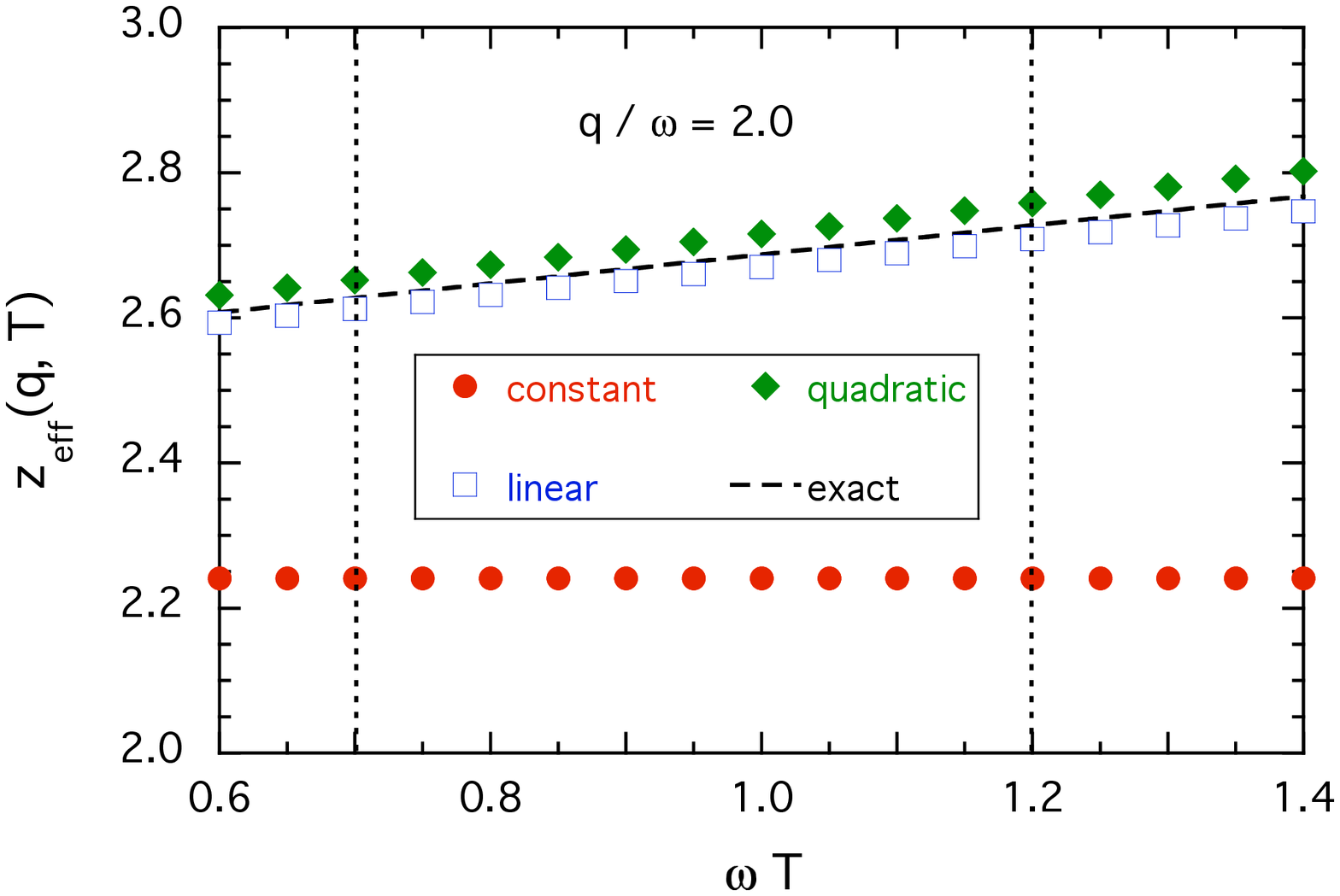}&
\includegraphics[width=5.5cm]{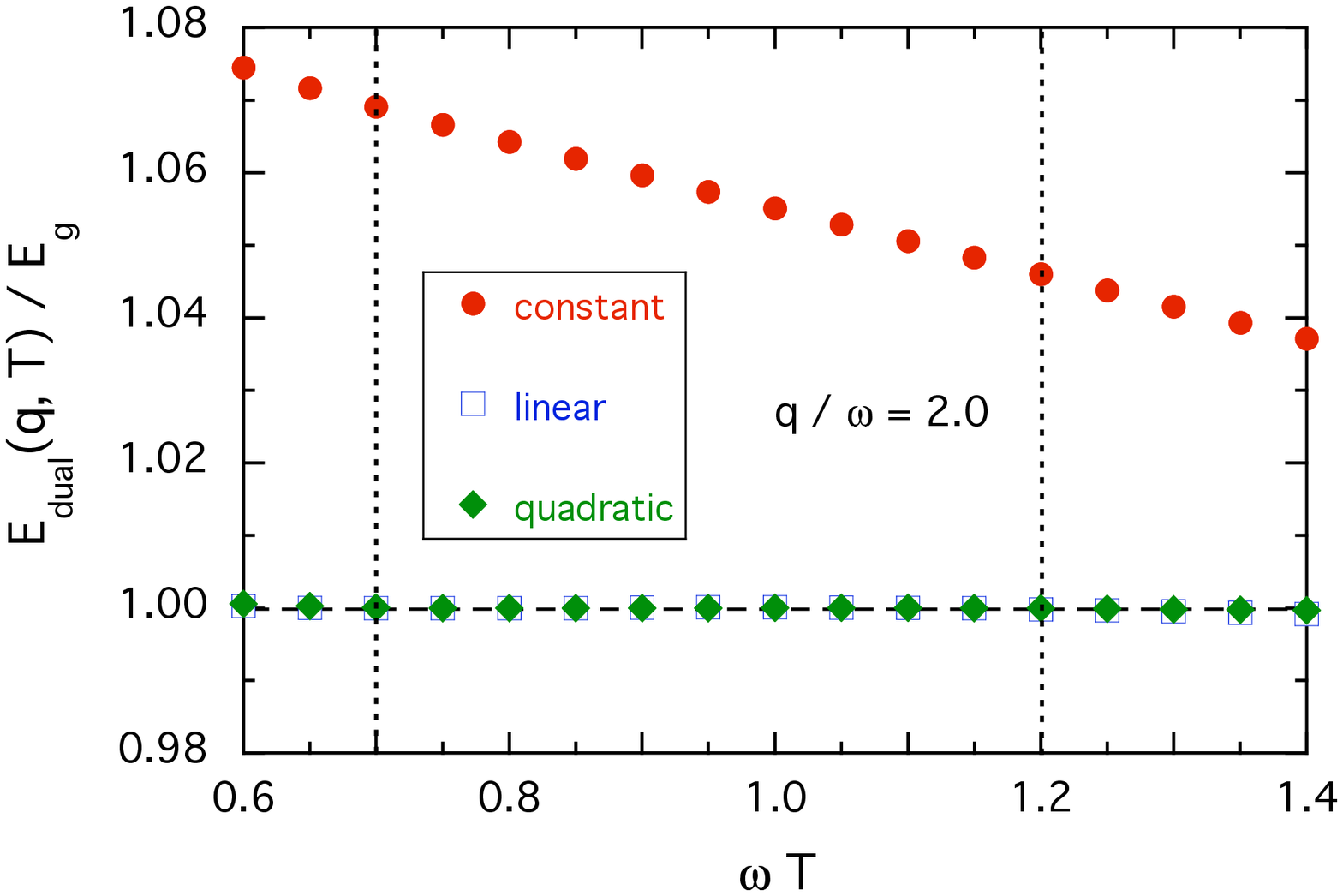}&
\includegraphics[width=5.5cm]{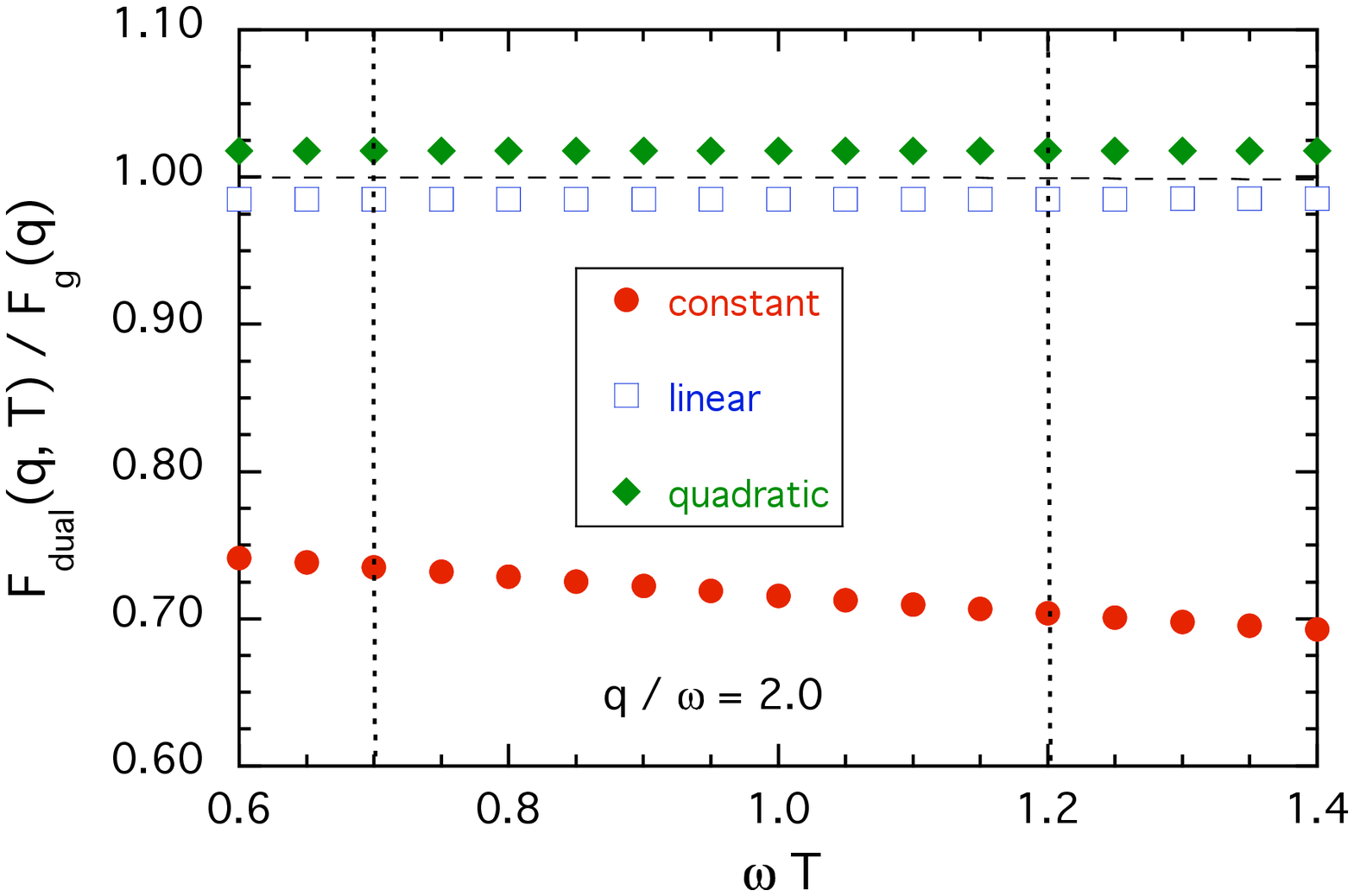}\\
\end{tabular}
\caption{\label{Plot:5} 
The extraction of the ground-state form factor at 
$q/\omega=0.5$ (first row),  
$q/\omega=1.5$ (second row), and $q/\omega=2$ (third row).  
First column: the effective continuum threshold as found by the fitting procedure (\ref{chisq}); 
second column: the fitted dual energy $E_{\rm dual}(T,q)/E_{\rm g}$; 
third column: the corresponding form factor ratio $F_{\rm dual}(T,q)/F_{\rm g}(q)$.
Dashed black line: exact $z_{\rm eff}$; 
full circles (red): constant $z^C_{\rm eff}$; empty squares (blue): linear $z^L_{\rm eff}(T)$; 
full diamonds (green): quadratic $z^Q_{\rm eff}(T)$. }
\end{figure}

\begin{figure}[!ht]
\includegraphics[width=10.5cm]{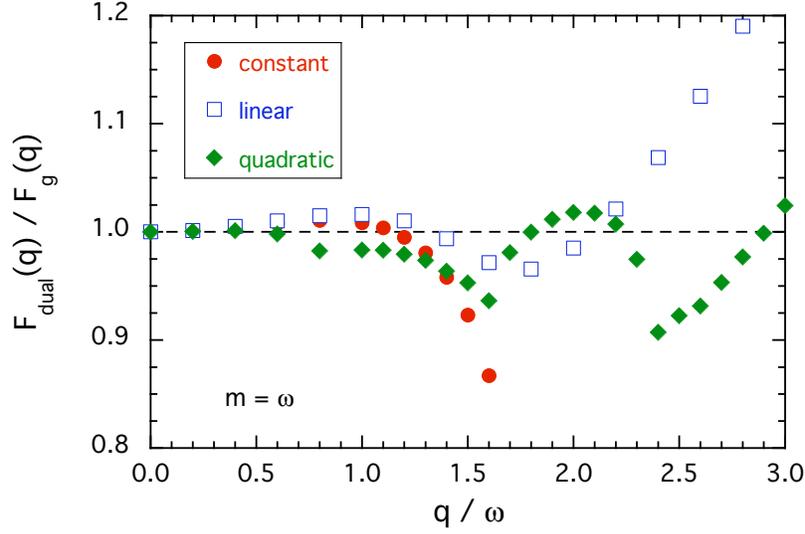}
\caption{\label{Plot:6}
The ratio $F_{\rm dual}(q)/F_{\rm g}(q)$ of the form factor $F(q)$ extracted from the sum rule (\ref{srgamma}), using 
different approximations for $z_{\rm eff}(T, q)$, and the exact 
ground-state form factor $F_{\rm g}(q)$, given by Eq.~(\ref{FG}), versus the 
momentum transfer $q$.
The dots, squares and diamonds correspond, respectively, to the results obtained using 
a constant (\ref{constant}), linear (\ref{linear}) and quadratic (\ref{quadratic}) 
approximation for the $T$-dependence of $z_{\rm eff}(T, q)$.}
\end{figure}

The results for the form factor $F(q)$ obtained from the dual correlator by optimizing 
the parameters of the three Ans\"atze (\ref{constant})--(\ref{quadratic}) according to (\ref{chisq}) 
are shown in Figs.~\ref{Plot:5} and \ref{Plot:6}. 
For a $T$-independent approximation (\ref{constant}), the minimization of $\chi^2$ (\ref{chisq}) 
leads to $z_0^C(q)$ for which the dual energy $E_{\rm dual}(T, q)$ differs from $E_{\rm g}$ by less than 
$1 \%$ and the dual form factor $F_{\rm dual}(q)$ turns out to be practically $T$-independent 
in the whole fiducial range. Such a stability, usually referred to as the Borel stability, 
is often (erroneously)  
claimed to be the way to control the accuracy of the extracted form factor. From Fig.~\ref{Plot:5} it 
is clear, however, that the actual error of the extracted form factor turns out to be much larger than 
the variation of the form factor in the fiducial range of $T$. 
Moreover, the T-independent Ansatz is applicable only for not very large $q$: for $q \gtrsim 1.7\, \omega$ 
it fails to reproduce the ground-state energy in the fiducial range.

Allowing for a linear Ansatz (\ref{linear}) for $z_{\rm eff}(T, q)$ extends the range 
of the momentum transfers where the form factor can be extracted from the sum rule with a reasonable accuracy. 
Note that the {\em exact} effective continuum threshold of the HO model can be well approximated 
by a linear function of $T$ in the whole fiducial range, see Fig.~\ref{Plot:3}(a). 
Nevertheless, this does not mean that the minimization of (\ref{chisq}) finds the {\em exact}  
effective continuum threshold and leads to the {\em exact} form factor: the 
deviation of the extracted form factor from the exact one amounts to a few percent for 
$q \lesssim 2 \omega$ and increases dramatically for $q \gtrsim 2\, \omega$, see Fig.~\ref{Plot:6}.
The dramatic increase of the error at $q \gtrsim 2\omega$ is directly related to the fact that the 
contribution of the ground state to the correlator decreases rapidly with $q$ in the given $T$-window 
$0.7\le \omega T\le 1.2$ (see Fig.~\ref{Plot:1}). 
Let us emphasize that the actual error of the extracted form factor turns out to be much larger than the variation 
of the form factor in the window, see Fig.~\ref{Plot:5}. 

One may try to go further and to consider the quadratic Ansatz (\ref{quadratic}).
From Fig.~\ref{Plot:6} it is clear that the quadratic Ansatz leads to certain 
instabilities in the extracted value of the form factor and does not, in general, improve the actual 
accuracy of the form factor extraction.
This is not strange: such a behaviour just reflects the fact that the sum-rule 
extraction of the ground-state parameters can be performed only with a limited accuracy, 
which has been known since the initial SVZ paper \cite{svz}. 
It should be therefore clear that there is no way to get further improvement in the 
accuracy of the extracted form factor by increasing the degree of the polynomial 
approximation of $z_{\rm eff}(T, q)$.

Nevertheless, we have seen a clear improvement in the outcome of the extraction procedure as one 
goes beyond the assumption of a $T$-independent effective continuum threshold: 
first, the actual accuracy turns out to be (much) better; 
second, the $T$-dependent Ansatz allows one to 
extract the form factor in a broader range of the momentum transfer. 

The crucial question for the application of sum rules to the bound-state form factors is how to understand 
the actual accuracy of the ground-state parameter extracted from the sum rule. 

\noindent 
(i) Obviously, the Borel stability does not guarantee the good extraction of the ground-state parameter 
and the variation of the extracted ground-state parameter in the window does not provide a realistic error estimate. 

\noindent 
(ii) It should be understood that even a very accurate reproduction of the ground-state energy 
in the window does not lead to equivalently accurate extraction of the ground-state parameters. 
This is rather unpleasant since the deviation of the dual energy from the 
known ground-state energy is the only visible characteristic which can be controlled in the 
realistic cases. What really matters for the extraction of the bound-state parameters is the 
deviation of the fitted effective threshold from the exact effective threshold which, however, 
in the realistic cases remains unknown. 

In the HO model, a realistic error estimate of the accuracy in a wide range of values of $q$ 
may be obtained by comparing to each other the form factor values extracted by assuming the linear 
and quadratic Ans\"atze for the effective continuum threshold (\ref{linear}) and (\ref{quadratic}). 
Whether this feature persists in QCD is an interesting and important issue to be addressed 
in the future.

\section{\label{Sect:Conclusions}Conclusions}
We studied the extraction of the ground-state form factor from the vacuum-to-vacuum correlator 
in the exactly solvable harmonic-oscillator model applying the standard procedures of the 
method of QCD sum rules. 
Let us summarize the main messages of our analysis:

\begin{itemize}
\item 
The knowledge of the correlator in a limited range of relatively small Euclidean 
times $T$ (that is, large Borel masses) is not sufficient for the determination of the
ground-state parameters. In addition to the OPE for the relevant correlator, one needs an 
independent criterion for fixing the effective continuum threshold. 

\item 
Assuming a $T$-independent (i.e., a Borel-parameter independent) effective 
continuum threshold, the error of the extracted form factor $F(T,q)$ 
turns out to be typically much larger than 
(i) the error of the description of the exact correlator $\Gamma$ by the truncated OPE $\Gamma_{\rm OPE}$ and 
(ii) the variation of $F(T,q)$ in the fiducial range of $T$ (i.e., the Borel window).
As the result, the actual value of the hadron form factor may lie far outside the range 
covered by the sum-rule estimate $F(T,q)$ within the Borel window of $T$. 
Therefore, the variation of the hadron form factor (as well as other hadron parameters) 
within the Borel window cannot be used as an estimate of its systematic error. 
The latter point is of particular relevance for the practical applications of sum rules in QCD, 
since the Borel stability is usually (erroneously) believed to control the accuracy and the reliability 
of the extracted ground-state parameter. 

\item 
In the cases where the ground-state mass is known (e.g., experimentally measured), the actual accuracy 
of the sum-rule 
analysis may be considerably improved by allowing for a $T$-dependent effective continuum threshold and  
finding its parameters by minimizing the deviation of the energy of the dual correlator from 
the known value of the ground-state energy, Eq.~(\ref{chisq}).  
  
In the HO model, this procedure was shown to yield clear improvements in the extracted values of the decay 
constant and the form factor. 

Moreover, in the HO model the deviation between the hadron parameter  
values obtained by using different Ans\"atze for $z_{\rm eff}$ gives {\it de facto} a realistic 
error estimate. This was observed for both the decay constant and the form factor.

\end{itemize}

Unfortunately, it remains impossible to construct the band of values which 
may be {\it proven} to contain the actual form factor. In this sense, the method of sum rules 
is not able to provide {\it rigorous} error estimates. Nevertheless, the application of the new  
procedures formulated in this paper to hadron form factors in QCD seems very promising with 
respect to improving the {\it actual} accuracy of the method. 

\vspace{1cm}

\noindent {\it Acknowledgments:} 
D.~M. gratefully acknowledges financial support from the Austrian
Science Fund (FWF) under project P20573 and from the President of 
Russian Federation under grant for leading scientific schools 1456.2008.2.


\begin{thebibliography}{30}
\bibitem{svz}
M.~Shifman, A.~Vainshtein, and V.~Zakharov, Nucl.~Phys.~B {\bf 147}, 385 (1979).
\bibitem{ioffe}
B.~L.~Ioffe and A.~V.~Smilga, Phys.~Lett.~B {\bf 114}, 353 (1982); 
V.~A.~Nesterenko and A.~V.~Radyushkin, Phys.~Lett.~B {\bf 115}, 410 (1982).
\bibitem{lms_lcsr}
W.~Lucha, D.~Melikhov, and S.~Simula, Phys.~Rev.~D {\bf 75}, 096002 (2007); 
Phys.~Atom.~Nucl.~{\bf 71}, 545 (2008).
\bibitem{lms_2ptsr}
W.~Lucha, D.~Melikhov, and S.~Simula, Phys.~Rev.~D {\bf 76}, 036002 (2007); 
Phys.~Lett.~B {\bf 657}, 148 (2007); \\
Phys.~Atom.~Nucl.~{\bf 71}, 1461 (2008).
\bibitem{lms_3ptsr}
W.~Lucha, D.~Melikhov, and S.~Simula, Phys.~Lett.~B {\bf 671} 445, (2009).
\bibitem{m_lcsr}
D.~Melikhov, Phys.~Lett.~B {\bf 671} 450, (2009).
\bibitem{lms_prl}
W.~Lucha, D.~Melikhov, and S.~Simula, arXiv:0902.4202 [hep-ph].
\bibitem{nsvz}
V.~Novikov, M.~Shifman, A.~Vainshtein, and V.~Zakharov, Nucl.~Phys.~B {\bf 237}, 525 (1984).
\bibitem{nsvz1}
A.~I.~Vainshtein, V.~I.~Zakharov, V.~A.~Novikov, and M.~A.~Shifman, Sov.~J.~Nucl.~Phys.~{\bf 32}, 840 (1980).
\bibitem{qmsr}
V.~A.~Novikov {\it et al.}, Phys.~Rep.~{\bf 41}, 1 (1978); 
M.~B.~Voloshin, Nucl.~Phys.~B {\bf 154}, 365 (1979);
J.~S.~Bell and R.~Bertlmann, Nucl.~Phys.~B {\bf 177}, 218 (1981); Nucl.~Phys.~B {\bf 187}, 285 (1981); 
V.~A.~Novikov, M.~A.~Shifman, A.~I.~Vainshtein, and V.~I.~Zakharov, Nucl.~Phys.~B {\bf 191}, 301 (1981).
\bibitem{orsay}
A.~Le Yaouanc {\it et al.}, Phys.~Rev.~D {\bf 62}, 074007 (2000); 
Phys.~Lett.~B {\bf 488}, 153 (2000); Phys.~Lett.~B {\bf 517}, 135 (2001).
\bibitem{ms_inclusive}
D.Melikhov, S.Simula, Phys.~Rev.~D{\bf 62}, 074012, 2000. 
\bibitem{radyushkin2001}A.~V.~Radyushkin, 
in {\it Strong Interactions at Low and Intermediate Energies}, edited by J.~L.~Goity, 
Singapore, World Scientific, pp. 91--150 (2000) [hep-ph/0101227].
\bibitem{bakulev}A.~P.~Bakulev, Acta Phys.~Polon.~B {\bf 37}, 3603 (2006) [hep-ph/0610266]. 
\bibitem{lms_prd75}
D.~Melikhov, Eur.~Phys.~J.~direct~C {\bf 4}, 2 (2002) [hep-ph/0110087]; 
D.~Melikhov and S.~Simula, Eur.~Phys.~J.~C {\bf 37}, 437 (2004); 
W.~Lucha, D.~Melikhov, and S.~Simula, Phys.~Rev.~D {\bf 75}, 016001 (2007).
\end{thebibliography}
\end{document}